\documentclass[aps,
 amsmath,amssymb,
 aps,
]{revtex4-2}
\usepackage{booktabs}
\usepackage{graphicx}
\usepackage{dcolumn}
\usepackage{bm}

\usepackage{braket}
\usepackage{ mathrsfs }
\usepackage{graphicx}
\graphicspath{{images/}}
\newcommand{\bey}{\begin{eqnarray}}
\newcommand{\eey}{\end{eqnarray}}

\newcommand{\R}{\mathbb{R}}

\begin{document}


\title{Beyond the Unruh vacuum: multi-time correlations in black hole collapse and evaporation }

\author{Konstantinos Xenos}
\email{constantinxenos@gmail.com}
 \author{Charis Anastopoulos}
 \email{anastop@upatras.gr}
\author{Andreas F. Terzis}
\email{afterzis@upatras.gr}

 \affiliation{Laboratory of Universe Sciences, Department of Physics, University of Patras, 26504 Patras, Greece}
 
 \begin{abstract}
The black hole information paradox originates from the thermal character of Hawking radiation, which appears to erase information about the collapsing matter. However, thermality constrains only observables defined at a single time and leaves the structure of temporal quantum correlations largely unexplored. Here we show that multi-time quantum-field correlations provide a concrete mechanism for the survival of pre-collapse information in black hole evaporation.
Using a two-dimensional model of gravitational collapse and evaporation, we demonstrate that late-time multi-time correlations are not fully reproduced by the Unruh vacuum. In particular, they contain a contribution that depends explicitly on parameters characterizing the pre-collapse state, despite the thermal character of the asymptotic radiation.
Our results identify measurable multi-time correlations as carriers of information in Hawking radiation and suggest that formulations of the black hole information paradox based solely on single-time observables are incomplete.

\end{abstract}
\maketitle

The black hole information paradox is commonly formulated in terms of asymptotic observables, whose late-time behavior is governed by the Hawking–Wald (HW) theorem \cite{Hawk1, Wald1}. This theorem implies that the reduced quantum state of fields at future null infinity $\mathscr I^{+}$ is asymptotically thermal at temperature $T_H$, so that correlations between asymptotic modes are entirely determined by the thermal state. In this framework, complete evaporation appears to map pure states to mixed ones, suggesting a loss of information. This conclusion underlies much of the modern discussion of the paradox, including proposals based on Page curves and firewall arguments \cite{Page93, AMPS, Mathur}, as well as viewpoints that question the necessity of unitarity in black hole evaporation \cite{Hawk76, UnWa}.

The HW theorem characterizes the reduced state on asymptotic hypersurfaces approaching $\mathscr I^{+}$, and therefore constrains observables associated with a single asymptotic time slice. However, physically measurable quantities in quantum field theory (QFT) are often multi-time correlations, which encode temporal coherences of the field. These are routinely probed in quantum optics, where they underlie phenomena such as photon bunching, antibunching, and time-resolved two-photon interference \cite{HBT, QOptics}. Analogous multi-time measurements arise in condensed matter systems, where current--current correlations and full counting statistics probe temporal correlations of electron transport \cite{NaBl}, and in ultracold atomic gases through time-resolved density--density correlations \cite{BDZ}. Such observables lie outside the scope of the HW theorem \cite{AnSav20}.

A proper treatment of such observables requires a measurement framework that assigns probabilities to spacetime-localized detection events. A systematic framework for multi-time observables in quantum field theory is provided by the Quantum Temporal Probabilities (QTP) program \cite{QTP4}, in which the probability density for $n$ detection events at spacetime points $x_i$ is a linear functional of the $2n$-point correlation function
\bey
G^{(n,n)}(x_1, \ldots, x_n; x_1', \ldots, x_n') := \mbox{Tr} \left\{{\cal T}\left[ \hat{O}(x_n) \ldots \hat{O}(x_1) \right] \hat{\rho}_0
\, \bar{\cal T}\left[ \hat{O}(x_1') \ldots \hat{O}(x_n') \right] \right\},
\label{nmpt}
\eey
where ${\cal T}$ and $\bar{\cal T}$ denote time- and anti-time-ordering, $\hat{O}(x)$ are local composite operators, and $\hat{\rho}_0$ is the initial state. The QTP formalism introduces no new dynamics; rather, it makes explicit features already encoded in field correlation functions. It is closely related to the Schwinger--Keldysh formalism in non-equilibrium QFT \cite{ctp1, ctp2, ctp3} and to Glauber's photodetection theory in quantum optics \cite{Glauber1, Glauber2}, providing a natural framework for temporally extended measurements.

In Ref.~\cite{AnSav20}, multi-time correlations in Hawking radiation for the Unruh vacuum of an eternal black hole were shown to exhibit strong non-thermal features, unrelated to grey-body factors \cite{Page2}, that retain memory of the emission history. This indicates that information can be encoded in higher-order temporal correlations of the quantum field.

Here we extend this analysis to a dynamical setting of gravitational collapse and evaporation. Using a two-dimensional model that allows exact evaluation of multi-time correlations, we show that late-time correlators depend explicitly on parameters characterizing the pre-collapse state.

Information is therefore not erased, but encoded in temporally extended correlations that remain accessible at late times.
These results identify a concrete mechanism by which deviations from thermality carry information within quantum field theory in curved spacetime, without invoking quantum gravity or nonlocal dynamics.

\bigskip

\section{Detector correlations in gravitational collapse}

We consider pointlike detectors that record energy while following prescribed spacetime trajectories specified as $x(t)$. Then, the QTP probability densities for single and double detection events are
\bey
P_1(t, E) &=& \int ds\, e^{-iE s} g(s)\,
G^{(1,1)}\!\left(x(t + \tfrac{s}{2}); x(t - \tfrac{s}{2})\right), \label{p1te} \\
P_2(t_1, E_1; t_2, E_2) &=& \int ds_1 ds_2\, e^{-iE_1 s_1 - iE_2 s_2} g(s_1) g(s_2) \nonumber \\
&\times& G^{(2,2)}\!\left(x_1(t_1 + \tfrac{s_1}{2}), x_2(t_2 + \tfrac{s_2}{2});
x_1(t_1 - \tfrac{s_1}{2}), x_2(t_2 - \tfrac{s_2}{2})\right), \label{p2te}
\eey
where $t_i$ are proper times along the detector trajectories and $g(s)$ encodes the detector response, typically decaying on a timescale $\sigma$. In the regime $E\sigma \gg 1$, one may approximate $g(s)\simeq g(0)$, corresponding to sharp energy resolution; measurable probabilities are obtained by additional smearing at the scale $\sigma$. 

\begin{figure}
\includegraphics[width=5cm]{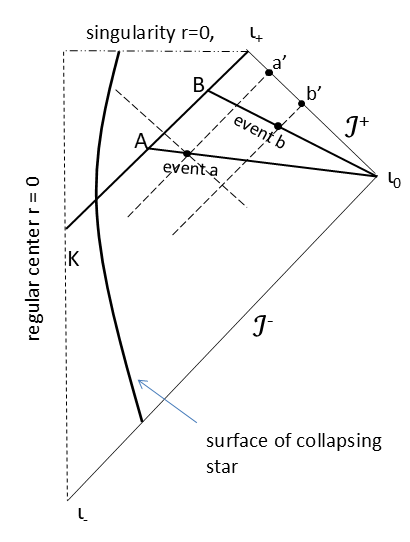}
\caption{
Penrose diagram of an evaporating black hole. Two measurement events $a$ and $b$ lie on distinct late-time Cauchy surfaces $KA\iota_0$ and $KB\iota_0$ and are spacelike separated, implying vanishing correlations in a local description. When projected onto future null infinity $\mathscr I^+$, the corresponding points $a'$ and $b'$ appear causally connected. The figure illustrates that projection to null infinity does not preserve the full spacetime relation between detection events, motivating a real-time treatment of correlations.
 }
\label{penrose}
 \end{figure}

The S-matrix formulation underlying the HW theorem characterizes information on asymptotic hypersurfaces approaching $\mathscr I^{+}$, and therefore does not fully encode correlations between events occurring at different times \cite{AnSav20}. As illustrated in Fig.~\ref{penrose}, projecting spacetime events onto null infinity can misrepresent their causal and correlational structure. A proper analysis of information retention in black hole evaporation therefore requires real-time formulations of QFT---such as the Schwinger--Keldysh framework---together with operational theories of quantum measurements \cite{HeKr, Sorkin, QTP1, OkOz, QTP3, FeVe, GGM22, PTM, QTP4}.

We will use the   probability densities (\ref{p1te}) and (\ref{p2te}) in order to analyze 
temporal correlations in Hawking radiation, and the possible retrieval of pre-collapse information from such correlations.  
 To this end, we 
employ the paradigmatic model of a collapsing null shell in a 1+1-dimensional spacetime \cite{Unruh76, BiDa}. The metric is 
\bey
ds^2 = \left\{ \begin{array}{cc} d\tau^2 - dr^2, & r \leq R(\tau) \\ 
\left(1-\dfrac{2M}{r}\right)\Big( dt^2 - dr^{*2}),& r> R(\tau) \end{array} \right.,
\eey
where $R(\tau)$ is the trajectory of the shell, parameterized in terms of the Minkowski time coordinate $\tau$ inside the shell;  the 
tortoise coordinate $r^*$ is defined as  $dr^* = dr/(1-2M/r)$.

The collapse begins at time $\tau=0$; we take $R(\tau) = R_0 > 2M$ for $\tau < 0$. The simplest collapse law is linear   $ R(\tau)=R_0 - \nu \tau$, with $  R_0 > 2 M$, the horizon forming at   $\tau_h= (R_0 -2M)/\nu$.

 We introduce the incoming and outgoing null coordinates in the shell's interior: $U = \tau-r+R_0$, $V= \tau + r -R_0$; and in the shell's exterior: $u = t-r^*+R^*_0$, $ v= t + r^* -R^*_0$. Then, $ds^2 = dU dV$ in the interior and $ds^2= (1-2M/r) \; du dv$ in the exterior. At the beginning of the collapse,   $u= U = v = V = 0$ on the shell. 

Next we consider   a quantum massless scalar field $\hat{\phi}$ obeying the Klein-Gordon equation $\;\square \hat{\phi} = 0$. We quantize the field with respect to the set of positive-frequency modes that are purely incoming at past null infinity $\mathscr I^{-}$. Taking Dirichlet boundary conditions at $r = 0$, we identify the modes \cite{Unruh76, BiDa} as 
\begin{equation}\label{modes}
\phi_\omega (u,v) =\frac{1}{\sqrt{4\pi \omega}} \left( \;e^{-i\omega v} - e^{-i\omega p(u)} \right). 
\end{equation}
The function $p(u) = U(u) - 2R_0$ can be implicitly evaluated for any collapse trajectory $R(\tau)$ \cite{BiDa}. In the Appendix, we show that for a linear collapse law with velocity $\nu = 1$, 
\bey
p(u) = -4M\left( W[D_0e^{-\frac{u}{4M}}] +1 \right),\;\;  D_0=\dfrac{U_h}{4M}e^{\frac{U_h}{4M}},
\eey
where $U_h\equiv U(\tau_h)= 2R_0-4M$ is the value of $U$ at horizon formation, and $W$ is the principal branch of the Lambert function---the inverse of the function $f(x) = xe^x$ \cite{Corless02}.  
 
The field operator is then expanded in this complete set of modes as :
\bey
    \hat\phi(u,v)=\sum_\omega \left( \hat{a}_\omega\; \phi_\omega(u,v)+\hat{a}_\omega^{\dagger} \; \phi_\omega^*(u,v) \right) \label{field quant}
\eey
with $\hat{a}_{\omega},\hat{a}^{\dagger}_{\omega}$ the standard annihilation and creation operators, obeying the usual commutation relations. The in-vacuum is defined by the condition $\hat{a}_{\omega}|0\rangle = 0$.

We assume an Unruh-DeWitt (dipole) coupling of the field to the detector \cite{Unruh76, Dewitt}, that is, we take the composite operators $\hat{O}(x)$ that define the field correlation functions to coincide with the scalar field $\hat{\phi}(x)$. Then, the correlation function $G^{(1,1)}$ in Eq. (\ref{p1te}) coincides with the Wightman function $ G^{(1,1)}(x;x') \equiv \langle 0 | \hat\phi(x)\,\hat\phi(x') | 0 \rangle 
  $, where $x=(u,v)$.
  
  We straightforwardly evaluate
\bey \label{2-point function}
      G^{(1,1)}(x;x')  = -\dfrac{1}{4\pi}\log\left(\dfrac{(v-v'-i\epsilon)\left( p(u)-p(u')-i\epsilon\right)}{(p(u)-v'-i\epsilon)\left( v-p(u')-i\epsilon\right)} \right).
\eey
with the  $i\epsilon$ prescription that ensures the correct analytic structure.
The Wightman function splits as $G^{(1,1)} = G_U^{(1,1)} - G_m^{(1,1)}$, where 
the term $G_U^{(1,1)}(x;x') = \log\left[(v-v'-i\epsilon)\left(U(u) - U(u')-i\epsilon\right)\right]$ coincides with the Wightman function for a scalar field in an eternal black hole spacetime at the Unruh vacuum. The only difference is that the Kruskal relation $U(u)$ is obtained at the limit of $u \gg 4M$. The remainder term $G_m^{(1,1)}(x;x') = \log\left[(p(u)-v'-i\epsilon)\left( v-p(u')-i\epsilon\right)\right]$ carries pre-collapse information encoded in the parameter $D_0$.

Unruh noticed that only the first term in the Wightman function   survives at large $v$. This implies that, after a transient period, Hawking radiation in a collapsing spacetime is indistinguishable from Hawking radiation 
in an eternal black hole, with the field at the Unruh vacuum. Indeed, by Eq. (\ref{p1te}), we straightforwardly calculate that the contribution from $G_m^{(1,1)}$ to the detection probability for a static detector is negligible. So, only the contribution from $G_U^{(1,1)}$ survives at late times, yielding the standard  result of a constant thermal detection rate: {$P_1(t, E) = \mathbf{Z}[E]/2E$}, where $\mathbf{Z}[E] = (e^{8\pi ME}-1)^{-1}$ is the Planck factor for the Hawking temperature.

\bigskip

\section{Information in multi-time correlations}
  
Unruh's  conclusion applies only to quantities defined at the level of two-point functions, like $P_1(t, E)$ above, or the expectation value of the stress-energy tensor. It does not apply to quantities defined at higher levels of correlation functions. 

To see this,   consider a dense network of detectors around a black hole, with one pair being active at each moment of time. The detection coincidences of Hawking quanta are described by a joint probability density 
$P_2(t_1, E_1; t_2, E_2)$, given by Eq. (\ref{p2te}). 
As in quantum optics, we define the second-order coherence function 
\bey
g^{(2)}(t_1, E_1; t_2, E_2) = \frac{P_2(t_1, E_1; t_2, E_2) }{P_1(t_1, E_1) P_1(t_2, E_2)}.
\eey
The evaluation of $g^{(2)}$ from Eq.~(\ref{p2te}) is technically lengthy but conceptually straightforward, and is presented in the Supplementary Information. It shows that $g^{(2)}$ can be expressed as a sum of three terms $ g^{(2)}= 1 + C^{(2)}_U +   C^{(2)}_{corr}+C^{(2)}_{mem}$, each with a different interpretation: the Unruh vacuum term $C^{(2)}_U$ which provides the   baseline; a memory term $C^{(2)}_{mem}$ that  carries pre-collapse information; and a non-Unruh correction $C^{(2)}_{corr}$ that captures additional deviations but not carrying any pre-collapse information.

The term $C^{(2)}_U$ is constructed solely from the Unruh-vacuum contribution to the two-point functions. For a pair of static detectors at  locations $r_1^*$ and $r_2^*$,
\begin{align}\label{cunruh}
     C^{(2)}_U(t_1, E_1; t_2, E_2)&=-\dfrac{E_1}{\mathbf{Z}^2[E_1]}\delta(E_1-E_2) F(\Delta t, E_1)  
\end{align}
where $\Delta t=t_2-t_1$. The exact form of the function $F(t, E)$ is given in the Appendix. Eq. (\ref{cunruh}) is analogous to the  expression computed in Ref. \cite{AnSav20} for an eternal black hole in four dimensions---their difference being due to the differences in the Wightman function between  two and four dimensions.

\begin{figure}[ht] 
  \centering  
  \includegraphics[width=12cm]{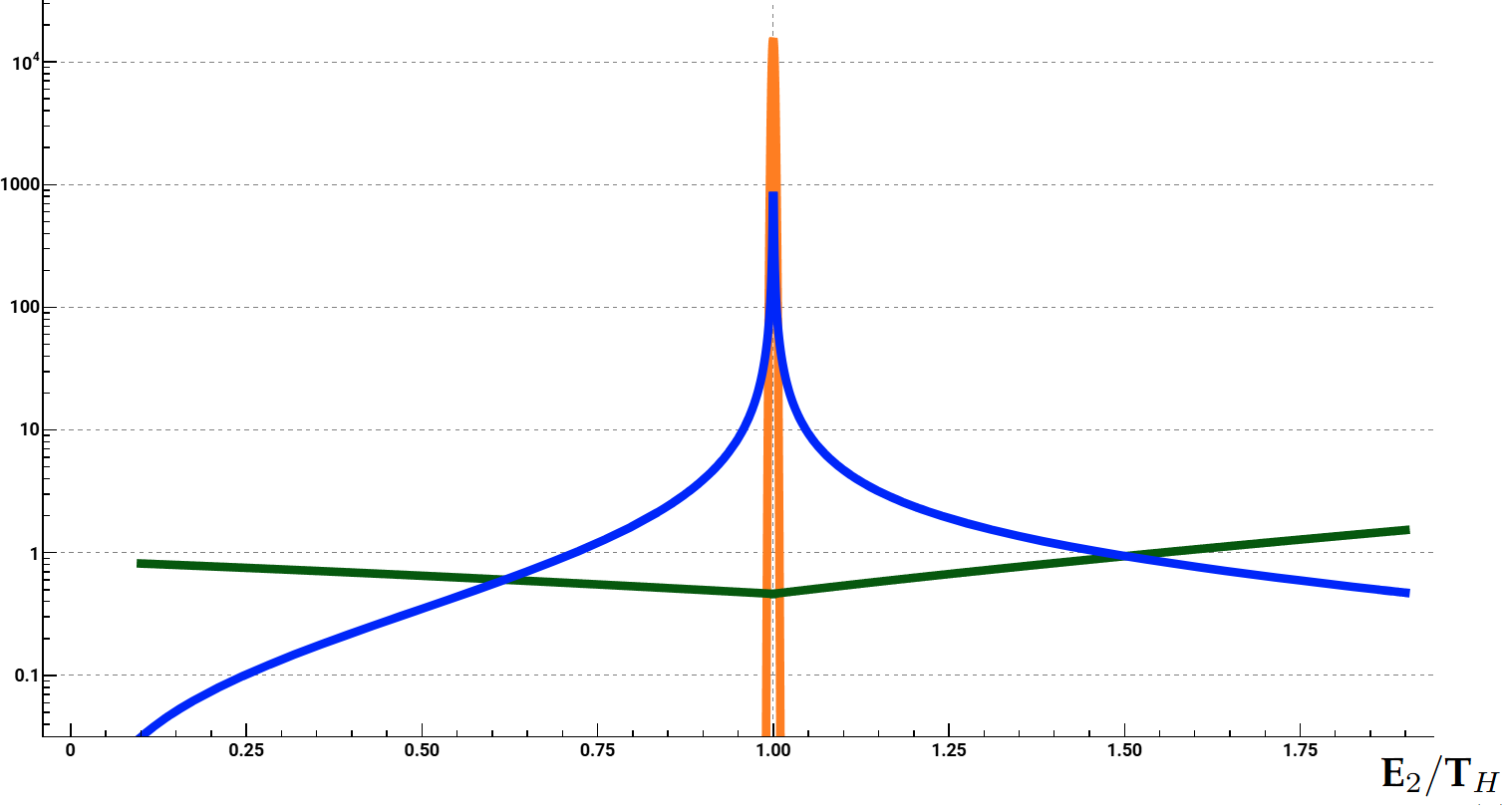}%
\raisebox{2.5cm}[0pt][0pt]{\includegraphics[width=0.15\textwidth]{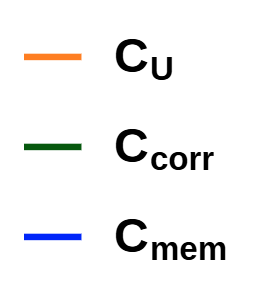}}
  \caption{The amplitudes of the three contributions $C^{(2)}_U$, $C^{(2)}_{corr}$  and $C^{(2)}_{mem}$ to the second-order coherence function for Hawking radiation are plotted as a function of $E_2/T_H$ for fixed $E_1 = T_H$, where $T_H = 1/8\pi M$ is the Hawking temperature. }
  \label{ERG}
\end{figure}

The correlations in Eq.~(\ref{cunruh}) are non-thermal, as they coincide with the correlations of a Gibbs state for the field only when the two detectors are co-located. As shown in Ref. \cite{AnSav11}, Unruh-type correlations decay exponentially with detector separation, whereas Gibbsian correlations exhibit a power-law fall-off. Nevertheless,  $C^{(2)}_U$ is proportional to $\delta(E_1 - E_2)$, like its Gibbsian counterpart, and therefore does not correlate photons of different energies.
The memory term $C^{(2)}_{mem}$ takes the form
\bey
    C^{(2)}_{mem}(t_1, E_1; t_2, E_2) = \dfrac{E_1}{(E_1-E_2)}\; \frac{\mathbf{Z}[2E_1]}{\mathbf{Z}[E_1]\mathbf{Z}[E_2]} \;\Theta(E_1-E_2) \cos(2E_1\Delta u+\phi_{2})\;\;+(1\leftrightarrow 2),
\eey
where  $\Delta u =  \Delta t + (r^*_1 - r^*_2)$, and  $(1\leftrightarrow 2)$ indicates the same terms with swapped indices for all  indexed variables;  $\Theta(z)$ is the Heaviside step function. Here, 
 \bey
\phi_{2} = 8M (E_1-E_2)\;\left[1 + \frac{v_2}{4M} + D_0e^{1 + \frac{\Delta t + r_1^*}{2M}}\right]. \label{phi2}
 \eey
The residual term $C^{(2)}_{corr}$ is well approximated by
\bey
C^{(2)}_{corr}(t_1, E_1; t_2, E_2) = \dfrac{\mathbf{Z}[2E_1]}{\mathbf{Z}[E_1]\mathbf{Z}[E_2]}\;   \;\Theta(E_2-E_1) \cos(2E_1\Delta u) \;\;+(1\leftrightarrow 2),
\eey
In Figure \ref{ERG}, we plot the amplitudes of the three correlation terms---that is, their maxima with respect to the oscillating phases.
 The memory-carrying contribution $C^{(2)}_{mem}$ is clearly distinguishable from both the Unruh term $C^{(2)}_U$ and the correction term $C^{(2)}_{corr}$.
Crucially, the
 non-Unruh terms encode correlations between photons of different energies. 
Such correlations exist neither in Gibbsian states nor in the Unruh vacuum. 
It is precisely in these terms that pre-collapse information resides: the contribution from $D_0$ vanishes when $E_1 = E_2$, and is therefore visible only in genuinely energy-resolved correlations.

All three terms contain rapidly oscillating phases with frequencies set by the detected energies, of order the Hawking temperature $T_H$. In realistic measurements with time resolution $\sigma$, the regime $T_H \sigma \gg 1$ suppresses these oscillations, leaving sharply peaked contributions at configurations where the phase vanishes.

The terms $C^{(2)}_U$ and $C^{(2)}_{corr}$ are peaked for pairs of events lying on the same lightcone, i.e., for $\Delta u = 0$ or $\Delta v = 0$. In contrast, the memory term $C^{(2)}_{mem}$ is peaked at
 $$2\Delta u + \frac{\phi_{2}}{E_1} = 0 \;\;\; \mbox{and} \;\;\;  2 \Delta u + \frac{\phi_{1}}{E_2}=0.$$

These conditions define a co-dimension one submanifold in the space of coincident detection events that extends also to late times.   Hence, the parameter $D_0$—the sole carrier of pre-collapse information in this model—appears directly in late-time correlations and is, in principle, observable. Crucially, the amplitude of the memory term is not suppressed (Fig.~\ref{ERG}): it is clearly distinguishable from the Unruh contribution and dominates over the residual term.

Although this information is, in principle, recoverable, its extraction is nontrivial. The memory contribution is intrinsically non-stationary, depending on the absolute detection times rather than their difference, which obstructs a straightforward statistical reconstruction of the corresponding correlators.

However, the information paradox concerns the existence of information, not the practical feasibility of extracting it. The dependence of the multi-time correlators on the parameter $D_0$ arises directly from the structure of quantum-field correlations and is therefore not tied to special features of the model. Our results thus provide explicit evidence that information about the pre-collapse state can survive in measurable late-time correlations within quantum field theory in curved spacetime, without invoking quantum gravity or nonlocal dynamics.

These findings establish a sharp distinction between single-time and multi-time observables in Hawking radiation. While the reduced state at $\mathscr I^+$ is thermal, multi-time correlations retain explicit dependence on the pre-collapse state. Any description based solely on single-time observables is therefore necessarily incomplete: information can remain present in temporally extended correlations even when absent from the asymptotic thermal state.

\section*{Acknowldgements}
K.X. acknowledges financial support from the Andreas Mentzelopoulos Foundation.
 C. A. acknowledges support from the COST Action CA23115 ``Relativistic Quantum Information".

\clearpage
\appendix
\section*{Appendix}
\addcontentsline{toc}{section}{Appendix} 

\section{The Hawking-Wald theorem and its limitations}
This section summarizes the results of Ref. \cite{AnSav19} about the limitations of the Hawking-Wald theorem.

  In the original analysis of black hole radiation, Hawking proved that particle numbers at the future null infinity $\mathcal{I}^+$ are characterized by a Planckian spectrum \cite{Hawk1}. Subsequently, Wald showed that {\em all} observables at $\mathcal{I}^+$ behave thermally \cite{Wald1}. We   refer to the latter result as the Hawking-Wald (HW) theorem. It is based on a scattering-matrix approach to quantum field theory (QFT) in curved spacetime. It implies that there is no correlation between different field modes of the emitted radiation \cite{Park75, Hawk76}.

  The HW theorem is commonly cited as a proof of the fact that no information can be stored in the correlations of Hawking radiation. However, the theorem refers only to asymptotic single-time properties of the quantum field, and it makes no  statement about multi-time measurements at late times. In this section, we present a  detailed analysis of the theorem that makes our point explicit.

Let $(M, g)$ be an asymptotically flat spacetime  that describes the collapse of a star leading to the formation of a black hole with future event horizon ${\cal H}^+$. Consider a free quantum scalar field $\hat{\Phi}(X)$ on $M$. To construct the  Hilbert space ${\cal F}$  of states for this field, one first identifies the real vector space $V$ of solutions to Klein-Gordon's (KG) equation with the KG inner product. Then, one complexifies $V$ in order to construct a complex Hilbert space $V_{\pmb C}$ \cite{QFTCSWald}. The Hilbert space ${\cal F}$ for the field degrees of freedom  is the exponential Hilbert space $e^{V_{\pmb C}}$, i.e., the bosonic Fock space associated to $V_{\pmb C}$
\begin{eqnarray}
{\cal F} = e^{V_{\pmb C}} := {\pmb C}\oplus V_{\pmb C} \oplus (V_{\pmb C} \otimes V_{\pmb C})_S \oplus  (V_{\pmb C} \otimes V_{\pmb C} \otimes V_{\pmb C} )_S \oplus \ldots,
\end{eqnarray}
where the index $S$ refers to symmetrization. To complexify $V$, one chooses a subset of complex-valued solutions $u_a(X)$  to the KG equation   to define an orthonormal basis on $V_{\pmb C}$.

In presence of a time-like Killing vector $\partial/\partial t$, the functions $u_a(X)$   are positive-frequency with respect to $t$,
 \begin{eqnarray}
 i \frac{\partial}{\partial t} u_a = \omega_a u_a, \label{posfrq}
 \end{eqnarray}
 for frequencies $\omega_a > 0$.

Once the family of solutions $u_a(X)$ has been chosen, the field operator is expressed as
\begin{eqnarray}
\hat{\phi}(X) = \sum_a \left[ \hat{a}_a u_a(X) + \hat{a}^{\dagger}_a u^*_a(X) \right],
\end{eqnarray}
in terms of the creation and annihilation operators on  the Fock space ${\cal F}$.
 
 Let $W$ be a closed linear subspace of $V_{\pmb C}$, and $W^{\bot}$ its complement. Then, the field Hilbert space splits as a tensor product \cite{Klau70}
 \begin{eqnarray}
 {\cal F}  = e^{W\oplus W^{\bot}} = e^{W} \otimes e^{W^{\bot}}. \label{split}
 \end{eqnarray}
Of particular relevance are tensor products of the form Eq. (\ref{split}) that are generated by partitioning a Cauchy surface.
Let $\Sigma$  be a Cauchy surface on $M$,  and  $C_1$, $C_2$ subsets of $\Sigma$, such that $C_1 \cap C_2 = \emptyset$ and $C_1 \cup C_2 = \Sigma$. We define a subspace $W_{C_1} $ of $V_{\pmb C}$ that consists of all functions $f(X) = \sum_a c_a u_a(X)$, such that $f(X) = 0$, for $X \in C_2$; $W_{C_2} = W_{C_1}^{\bot}$ is spanned by all functions $g(X)$ that vanish for $X \in C_1$.
The Hilbert space ${\cal H}_{C_i} = e^{W_{C_i}}$ describes field state localized in $C_i$, $i = 1, 2$. Hence, Eq. (\ref{split})  can be expressed
\begin{eqnarray}
{\cal F} = {\cal H}_{C_1} \otimes {\cal H}_{C_2}. \label{split2}
\end{eqnarray}

 We choose a basis $f_i(X)$ of solutions in $W$, such that  $f_i(X) = 0$ for $X \in C_2$, and a basis $g_j (X)$of solutions in $W^{\bot}$, such that  $g_j(X) = 0$ for $X \in C_1$. Then, the field operator can be written as
 \begin{eqnarray}
\hat{\phi}(X) = \sum_i \left[ \hat{a}^{(1)}_i f_i(X) + \hat{a}^{(1)\dagger}_i f^*_i(X) \right] + \sum_j \left[ \hat{a}^{(2)}_j g_j(X) + \hat{a}^{(2)\dagger}_j g^*_j(X) \right]
\end{eqnarray}
 where $\hat{a}^{(1)}$ and   $\hat{a}^{(2)}$ are annihilation operators restricted to the subspace $W$ and $W^{\bot}$ respectively.

Any state $|\Psi\rangle \in {\cal F}$ can be expressed as a linear combination $\sum_{A, B} \lambda_{AB} |A\rangle_{1}\otimes |B\rangle_{2}$, where $|A\rangle_1 $ and $|B\rangle_2 $ define orthonormal sets in $e^{W}$ and $e^{W^{\bot}}$, respectively. Any vector $|A\rangle_1$ can be constructed from the consecutive  action of creation operators $\hat{a}^{(1)\dagger}_i$ on a reference state $|0 \rangle_1$, and any vector $|B\rangle$ can be constructed from the consecutive  action of creation operators $\hat{a}^{(2)\dagger}_j$ on a reference state $|0 \rangle_2$.

Consider now a single-time measurement localized in the region $C_1$.  Suppose that the field interacts with a measuring apparatus through a composite operator
  $\hat{O}(X)$ that is a local functional of the field $\hat{\phi}(X)$. The interaction  depends only on the operators   $\hat{a^{(1)}}_i$ and $\hat{a}^{(1)\dagger}_i$; hence, it affects only the vectors $|A\rangle_{1}$. Therefore, all information about such measurements is contained in the reduced density matrix $\hat{\rho}_1$ on ${\cal H}_{C_1}$,
\begin{eqnarray}
{}_1\langle A|\hat{\rho}_1|B\rangle_1 := \sum_C \lambda_{AC}\lambda^*_{BC}. \label{rho1red}
\end{eqnarray}

In a collapsing black hole spacetime, we consider the Cauchy surface $\Sigma_A := (KA)\cup(A \iota_0)$ of Fig. 1. Let $C_1$ be its segment outside the black hole. After the end of the collapse, there   is a
  time-like Killing vector $\frac{\partial}{\partial t}$ outside the black hole. Hence,  the modes   $f_i(X)$ can be chosen to be positive-frequency with respect to $\frac{\partial}{\partial t}$; the corresponding frequencies $\omega_i$ are then interpreted as single-particle energies. Therefore, we can choose the  
basis $|A\rangle_1$ to be eigenstates of the particle number operators, i.e., the elements of the basis are labeled by   a sequence of particle numbers
 $\{n_i\}$ for each mode $i$.

 The key point is that the Hilbert space split above is relevant only for a single moment of time, i.e., a specific Cauchy surface $\Sigma$. Consider a different Cauchy surface $\Sigma'$, with a different split $C_1'$ and $C_2'$.  The functions $g_j(X)$ that vanish in $C_1$ do not, in general, vanish in $C_1'$\footnote[3]{The set of solutions to the KG equation that vanishes on both $C_1$ and $C_1'$ is   of measure zero in $V$.}. It follows that the field operator $\hat{\phi}(X)$ for $X \in C_1'$ does not vanish on $C_2$. A localized measurement at $X \in C_1'$ involves also the operators $\hat{a}^{(2)}_j$ and $\hat{a}^{(2)\dagger}_j$. The probabilities for  two-time measurement localized at $X \in C_1$ and $X' \in C_1'$ cannot be expressed solely in terms of the reduced density matrix, either of the ${\cal H}_{C_1} \otimes {\cal H}_{C_2}$ or of the ${\cal H}_{C'_1} \otimes {\cal H}_{C'_2}$ partition\footnote[4]{Since the Fock space split is time-dependent,  entanglement between modes outside the black hole and on the horizon is also time-dependent. Common statements about this entanglement  refer to its asymptotic value. It is far from obvious that this asymptotic expression remains relevant after the inclusion of backreaction. The effects of backreaction are not asymptotic, for example, the change in the black hole mass is manifested at finite times. However, the Hilbert space split (\ref{split2})  is   not unique at finite times. Any choice of $C_1$ and $C_2$, such that $C_1 \rightarrow  \mathcal{I}^+$ and $ C_2 \rightarrow {\cal H}^+$ leads to the same asymptotic value of entanglement. In our opinion, this is an indication that entanglement may not be the most appropriate measure of the correlations in Hawking radiation, especially in relation to black hole evaporation. One should look for a measure that incorporates information about multi-time correlations and it is uniquely defined at finite times.
  }. 

 The HW theorem  involves a split of the form Eq. (\ref{split}) in relation to the Cauchy surface $\Sigma_{\infty} = \mathcal{I}^+ \cup {\cal H}^+$, and it demonstrates that   the reduced density matrix at $\mathcal{I}^+$ is Gibbsian,
 \begin{eqnarray}
 \hat{\rho}_1 = \frac{1}{Z} \sum_{\{n_i\}} e^{-\sum_i n_i \omega_i/T_H} |\{n_i\}\rangle \langle \{n_i\}| \label{hawkrad}
 \end{eqnarray}
where $T_H = (8\pi M)^{-1}$ is the Hawking temperature, and $Z = \sum_{\{n_i\}} e^{-\sum_i n_i \omega_i/T_H}$ is the partition function. The thermal density matrix (\ref{hawkrad}) is approximate as it ignores transient effects, i.e., particles created during collapse. It also assumes unit transmission probability for all field modes under consideration, i.e.,  that all "emitted" particles  reach  $\mathcal{I}^+$.

The Cauchy surface $\Sigma_{\infty}$ is the limit of the Cauchy surface $\Sigma_A$ of Fig. 1 as $A \rightarrow \iota^+$. Hence, the HW theorem can be viewed as a statement about the asymptotic form of the reduced density matrix defined on the subset $C_1$ of $\Sigma_A$.  By construction, its conclusions are restricted to the outcomes of single-time measurements. The fact that multi-time measurements cannot be solely expressed in terms of a reduced density matrix is not affected by taking  one of the Cauchy surfaces to infinity.

For example,  two-time correlation may be   expressed in terms of the joint probability of detecting Hawking quanta by two apparatuses at two spacetime points in the Cauchy surfaces $\Sigma_A$ and $\Sigma_B$, as in the right-hand diagram of Fig. 1. The two detection events can have any separation  (timelike, spacelike, or null). They cannot be mapped to events on $\mathcal{I}^+$ without losing the key information of their causal relation.  

We conclude that the probabilities of
multi-time  measurements  cannot be expressed solely in terms of the Gibbsian reduced density matrix (\ref{hawkrad}).
Hence, multi-time correlations in Hawking radiation are generically non-thermal, in the sense that they do not coincide with correlations obtained from a Gibbsian state.

Next, we present  a more general argument why the HW theorem does not constrain multi-time correlations, based on well known properties of {\em quantum open systems} \cite{Dav, BrePe}.

The HW theorem focuses on field properties at the future null infinity $\mathcal{I}^+$.
Of course, no physical measurements occur literally at the null infinity. The HW theorem is best viewed as a statement about the long time limit of an {\em open quantum system}.  Consider the Cauchy surface $\Sigma_A := (KA)\cup(A \iota_0)$ of Fig. 1, and the reduced density matrix $\hat{\rho}_A$ obtained by tracing out the degrees of freedom of the surface $KA$. The field degrees of freedom (dofs) on the surface  $A \iota_0$ are effectively an open system, with the dofs at the horizon playing the role of the environment.  Hence, the HW theorem is a  statement about  asymptotic thermalization   in an open quantum system.

 The key point is that in open quantum systems, the time-evolving reduced density matrix of a subsystem  does not contain all information about the subsystem. It contains only information accessible by single-time measurements. It does {\em not} contain sufficient information to correctly reproduce the probabilities of multi-time measurements, {\em unless the open system dynamics is Markovian} \cite{PaZu93, VA17}.  

In non-Markovian processes, the environment keeps memory of properties of the system and releases this information to the system in a way that is not fully predictable by the open system dynamics.   {\em At the fundamental level},  open quantum systems that are defined by tracing out an environment have  non-Markovian dynamics. Markovian behavior emerges as a result of approximations. Hence, the HW theorem  rules out asymptotic single-time correlations in the Hawking radiation. However, it does not rule out  temporal ones, i.e., correlations defined in terms of   multi-time measurements.

\section{Elementary derivation of the QTP probability formula}

This section provides a brief summary of the QTP approach to quantum field measurements, and presents a simple derivation of its basic probability formula.

In the QTP approach,   the apparatus is fully incorporated into the quantum description and it is also treated via QFT. The interaction between the measured system and the apparatus local and causal, in the sense that it is governed by an interaction Hamiltonian that is a local functional of quantum fields.

The measurement apparatus is a macroscopic system that exhibits classical behavior: the pointer variable is a highly coarse-grained observable that satisfies specific decoherence conditions.

QTP treats measurements events as localized in space and in time, as is the case for all physical measurement outcomes. For example, consider a solid-state detector that is elementary in the sense that can record only a single event. The detector has a fixed location in a lab, and it records an event at a moment of time that  is determined with finite accuracy. In principle, both position and time can be random variables.  When directing a single particle towards an array of elementary detectors, both the specific  detector that records the particle (i.e., the locus of the record) and the time of recording vary from one run of the experiments to the other. Hence, physical predictions are expressed in terms of probability densities
        \bey
        P(x_1, q_1; x_2, q_2, \ldots, x_n, q_n), \label{probdengen}
        \eey
for multiple detection events. In Eq. (\ref{probdengen}), $x_i$ stand for spacetime points, $q_i$ stand for any other recorded observable and $P$ is a probability density with respect to both $x_i$ and $q_i$.

Consider a QFT on Minkowski spacetime $M$, with Heisenberg-picture fields $\hat{\phi}_r(x)$ defined on a Hilbert space ${\cal F}$
that carries a unitary representation of the Poincar\'e group.  The index $r$  runs over spacetime and internal indices.

We denote the Hilbert space associated to an apparatus by   ${\cal K}$. We assume that the apparatus follows a world tube ${\cal W} = \R \times U$ in Minkowski spacetime, for some $U \subset \R^3$.
We assume that the size of the apparatus is   much larger than the scale of microscopic dynamics.  

For a simple derivation of the QTP probability formula, 
we assume a field-apparatus coupling with support in a small spacetime region around a point $x$.  
The finite spacetime extent of the interaction  mimics the effect of a detection record localized at $x$. Working in the interaction picture, we express the coupling term as
\bey
\hat{V}_x = \int F_x(y) \hat{C}(y) \otimes \hat{J}(y), \label{VX}
\eey
where $\hat{C}(x)$ is a composite operator on ${\cal F}$ that is local with respect to the field $\hat{\phi}_r(x)$. The current operator  $\hat{J}^a(x)$ is defined on ${\cal K}$.
 The switching functions $F_x(y)$ are dimensionless. They vanish outside the interaction region and they depend on the motion of the apparatus. The spacetime volume  associated to a switching function is $\upsilon = \int dY F^2_x(y)$. 
 
 Note that the introduction of switching functions in the dynamics is an oversimplification. In the proper QTP derivation, the functions $F_x(y)$ describe the sampling of observables, and the interaction terms are Poincar\'e covariant.

The S-matrix associated to Eq. (\ref{VX}) is $\hat{S}_x = {\cal T} \exp[ - i \int d^4y F_x(y) \hat{C}(y) \otimes \hat{J}(y)]$, where ${\cal T}$ stands for time ordering. To leading order in the interaction,
\bey
\hat{S}_x = \hat{I} - i \hat{V}_x.
\eey
Let the initial state of the system  be $|\psi\rangle \in {\cal F}$ and the initial state of the apparatus  be $|\Omega\rangle$. A particle record appears if the detector transitions from $|\Omega\rangle$ to its complementary subspace ${\cal K}'$. Furthermore, on ${\cal K}'$, we  measure a property of the particle through a pointer observable $q$. The latter is described by a family of positive operators $\hat{\Pi}(q)$, such that $\sum_{q} \hat{\Pi}(q) = \hat{I} - |\Omega \rangle \langle \Omega|$. The pointer observable is typically very coarse, and we assume that it is stationary under spacetime translations by the self-dynamics of the detector, so that the record is preserved after the end of the measurement.

Then, we evaluate  the probability
\bey
\mbox{Prob}(x, q) = \langle \psi, \Omega| \hat{S}^{\dagger}_x[\hat{I} \otimes \hat{\Pi}(q)]\hat{S}_x|\psi, \Omega\rangle
\eey
 that the detector is excited and records a value $q$ to leading order in perturbation theory
\bey
\mbox{Prob}(x, q) = \int d^4y_1 d^4y_2  F_x(y_1) F_x(y_2) G_{ab}(y_1, y_2) \langle \Omega|\hat{J}^a(y_1) \hat{\Pi}(q) \hat{J}^b(y_2)|\Omega\rangle, \label{probX}
\eey
where
\bey
G_{ab}(x, x')  = \langle \psi|\hat{C}_a(x) \hat{C}_b(x')|\psi\rangle,
\eey
is a correlation function for the composite operator.

The probability $\mbox{Prob}(x, q)$ of Eq. (\ref{probX}) is not a density with respect to $x$, because $x$ appears as a parameter of the switching function.   We define an unnormalized probability density $W(x, q)$ with respect to $x$ by dividing $\mbox{Prob}(x, q)$ with the effective spacetime volume $\upsilon$,
\bey
W(x, q) = \upsilon^{-1}\mbox{Prob}(x, q). \label{Prob0b}
\eey

Some assumptions about the detector . First, we assume that the detector carries a representation of the spacetime translation group with generators $\hat{p}^{\mu}$.  At a fundamental level,  the detector Hilbert space also carries a representation of the Poincar\'e group. However, the state $|\Omega\rangle$ is not the Poincar\'e invariant vacuum.

We choose a reference point $x_0 $ in  the detector's world-tube, and we write
\bey
\hat{J}^{a}(y) = e^{-i \hat{p} \cdot (y - x_0)} \hat{J}^a(x_0) e^{i \hat{p} \cdot (y - x_0)}.
\eey
It is convenient to take
$|\Omega\rangle$ to be   {\em approximately translation invariant}. Intuitively, this corresponds to the idea that the apparatus is prepared in an initial state that is homogeneous at the length scales that correspond to position sampling and approximately static at the time scales that correspond to time sampling. In the present context, approximate translation invariance is the requirement that
\bey
\int d^4 x F_{x}(x') \hat{J}^a(x') |\Omega\rangle \simeq \int d^4 x F_{x}(x') e^{-i \hat{p} \cdot (x' - x_0)} \hat{J}^a(x_0)|\Omega\rangle.
\eey
With this assumption, we can write $\langle \Omega|\hat{J}^a(y_1) \hat{\Pi}(q) \hat{J}^b(y_2)|\Omega\rangle = R^{ab}(y_2 - y_1, q)$,
 where
\bey
R^{ab}(x, q) := \langle \Omega| \hat{J}^a(x_0) \sqrt{\hat{\Pi}}(q) e^{-i\hat{p}\cdot x}\sqrt{\hat{\Pi}}(q)\hat{J}^b(x_0)|\Omega\rangle \label{detkern}
\eey
We will refer to $R^{ab}(y, q)$ as the {\em detector kernel}.

The simplest choice for the switching functions $F_x$ are Gaussians,
\bey
F_x(y) = \exp[ - \frac{1}{2} D(x, y)],
\eey
where $D$ a Euclidean distance function on Minkowski spacetime. Such functions are defined in terms of a dimensionless Euclidean metric $g^{(E)}_{\mu \nu}$ on Minkowski spacetime. A simple and physically relevant metric is determined by the timelike vector field $u_{\mu}$ normal along the world-tube, by
\bey
g_{\mu \nu} = \frac{1}{\delta_t^2} u_{\mu} u_{\nu} + \frac{1}{\delta_x^2} ( u_{\mu} u_{\nu}  + \eta_{\mu \nu}).
\eey
where $\delta_t$ is the  temporal accuracy and $\delta_x$ is the special accuracy of the detector. As these quantities correspond to the sampling of the detection event, they are both macroscopic scales.

 Gaussian switching functions satisfy the identity
\bey
f(x) f(x') = f^2\left(\frac{x+x'}{2}\right) \sqrt{f}(x - x'). \label{gauidty}
\eey
The spacetime volume $\upsilon$ of the interaction region  is $\upsilon =\pi^2 \delta_t \delta_x^3$. We note that the function   $\sigma(x): = \frac{1}{\upsilon} f^2(x)$ is a normalized probability density on $M$. Then, we write
\bey
W(x, q) = \int d^4x'  \sigma(x - x') P(x', q), \label{probX20}
\eey
where
\bey
P(x, q) = \int d^4y  \sqrt{f}(y) R^{ab}(y, q) G_{ab}(x - \frac{1}{2}y, x +\frac{1}{2}y), \label{probX2}
 \eey


The probability distribution $W(x, q) $ is the convolution of $P(x, q)$ with the probability density $\sigma(x)$ that incorporates the accuracy of our measurements. If $P(x, q)$ is non-negative and the scale of variation in $x$ is much larger than both $\delta_t$ and $\delta_x$, we can treat $P(x, q)$  as a finer-grained version of $W(x, q)$ and employ this as our probability density for detection.

The kernel $R^{ab}(x, q)$ is typically characterized by some correlation length-scale $\ell$ and some correlation time-scale $\tau$, such that $ R^{ab}(x, q)  \simeq 0$ if $|t(\xi)| \gg \tau $ or  $|{\bf x}(\xi)|\gg \ell$.
Both scales $\ell$ and $\tau$ are microscopic and characterize the constituents of the apparatus and their dynamics. If $\ell \ll \delta_x$ and $\tau \ll \delta_t$, then $R^{ab}(x, q) \sqrt{f}(x) \simeq R^{ab}(x, q)$ and we obtain an expression for the probability density  $P(x, q)$ that is sampling-independent
\bey
P(x, q) = \int d^4 y     R^{ab}(y, q) G_{ab}(x - \frac{1}{2}y, x +\frac{1}{2}y). \label{prob1aa}
\eey
The probability densities (\ref{prob1aa}) are not normalized to unity. In general, the total probability of detection $P_{det} = \sum_{q} \int_{\cal W} d^4x  P(q, x) $ must be a small number, for perturbation theory to be applicable. There is always a probability $P(\emptyset) = 1 - P_{det}$ of no detection. We normalize probabilities by dividing $P(x, q)/P_{det}$, i.e., by conditioning the probability densities $P(q, x)$ with respect to the existence of a detection record.

\subsection{Remarks}

\noindent 1. The  definition (\ref{Prob0b}) of a spacetime density with respect to time is fully justified in classical probability theory, but it is not rigorous for quantum probabilities. There reason is that it involves a combination of probabilities defined with respect to different experimental set-ups, i.e., different switching functions for the Hamiltonians.
Nonetheless,   Eq. (\ref{Prob0b}) can  be derived as a genuine probability density in the context of the QTP method \cite{QTP3, QTPdet}, to leading order in the field-apparatus coupling. This derivation is reproduced in the Appendix.

\smallskip

\noindent 2. Probabilities are defined here using the Born rule for the pointer variable. Note that, strictly speaking, the probabilities  in  von Neumann measurements are defined at a time {\em after} the function has been switched off, and not at the time when the switching is on.
The  derivation in the appendix employs a  probability assignment for histories which incorporates both the Born rule and the state reduction rule.

\smallskip

\noindent 3.    In the proper QTP derivation of detection probabilities, the interaction is present at all times, as the total description must be time-translation invariant.
The smearing functions $F_x(y)$ are not interpreted in terms of a switching-on of the interaction,
 but they describe the {\em sampling} of the spacetime point. Hence, the spacetime volume $\hat{\upsilon}$ is a measure of {\em coarse-graining}, i.e., of the inaccuracy in the determination of the spacetime point. This point is important for a proper derivation, because probabilities can only be defined   for histories that satisfy a decoherence condition, for which coarse-graining is a prerequisite. In principle, a preferred value of $\upsilon$ that corresponds to the coarse-graining scale at which probabilities are well-defined is determined from first-principles---see \cite{QTP1} for explicit calculations in simple models.
  This means that not all sampling functions $F_x(y)$ are acceptable:  their support  cannot be made arbitrarily small. Such constraints cannot be seen in the derivation that we presented here.

\smallskip

\noindent 4. QTP leads to different predictions beyond the lowest order in the system-apparatus interaction. However, in many set-ups only the lowest order  terms can be said to provide a meaningful signal, i.e., a correlation between  observables in the measured field and pointer variables in the apparatus. Higher order interaction terms often degrade such correlations, and for this reason they can be conceptualized as noise. This  is the case, for example, in Glauber's photodetection theory.

\smallskip

\noindent 5. In general, the field-apparatus couplings  are fixed by the standard model of particle physics. Even in idealized models for detectors, we have to specify the physical process through which detection takes place. This information is encoded in  the composite operators $\hat{C}(x)$.
Consider, for example, the case of a QFT with a single scalar field $\hat{\phi}(x)$. The choice $\hat{C}(x) = \hat{\phi}(x)$ corresponds to detection of particles through absorption. The choice $\hat{C}(x) = :\hat{\phi}^2(x):$ corresponds to particle detection through scattering, through terms that involve one creation and one annihilation operator. This observation generalizes to fields of arbitrary spin.

\subsection{Detection probability for multiple detectors}
We proceed to the derivation of a probability formula for the case  that the field interacts with $n$ detectors. Each detector corresponds to a Hilbert space ${\cal K}_i$, $ i = 1, 2, \ldots, n$. The total Hilbert space of the system is ${\cal F} \otimes {\cal K}_1 \otimes {\cal K}_2 \otimes \ldots \otimes {\cal K}_n$. The coupling operator of the $i$-th detector is
\bey
\hat{V}^{(i)}_x = \int d^4y F_x(y) \hat{C}^{(i)}_a(y) \otimes \hat{I} \otimes \ldots \hat{J}_{(i)}^a(y) \otimes \ldots \otimes \hat{I}, \label{VXn}
\eey
where $\hat{J}_{(i)}^a(y)$ is a current operator for the $i$-th detector and $\hat{C}_a^{(i)}$ is the associated composite operator.

 We denote the S-matrix for the $i$-th detector by $\hat{S}^{(i)}_x$. Again, to leading order in the interaction $\hat{S}^{(i)}_x = \hat{I} - i \hat{V}^{(i)}_x $.
 We also denote the initial state of the $i$-th detector by $|\Omega_i\rangle$, and the measurement operators as $\hat{\Pi}^{(i)}(q_i)$.

The key point is that the S matrix for the total interaction of the field  is defined by time ordering with respect to the spacetime points  $x_1, x_2, \ldots, x_n$ associated to the detectors,
\bey
\hat{S}_{x_1, x_2, \ldots, x_n} = {\cal T}\left[ \hat{S}^{(1)}_{x_1} \hat{S}^{(2)}_{x_2} \ldots \hat{S}^{(n)}_{x_n}   \right];
\eey
here, ${\cal T}$ stands for time ordering.

To leading order in perturbation theory,  the probability density for $n$ measurement events is
\bey
W_n(x_1, q_1; x_2, q_2; \ldots; x_n, q_n) = \int d^4x'_1 \ldots d^4 x'_n \sigma^{(1)}(x_1 - x'_1) \ldots \sigma^{(n)}(x_n - x'_n)\nonumber \\ P(x'_1, q_1;  x'_2, q_2; \ldots; x'_n, q_n)
\eey
where
\bey
P_n(x_1, q_1; x_2, q_2; \ldots; x_n, q_n) = \int d^4 y_1 \ldots d^4 y_n   \sqrt{f^{(1)}}(y_1) \ldots \sqrt{f^{(n)}}(y_n)
R_{(1)}^{a_1b_1}(y_1, q_1) \ldots
\nonumber \\
\times \ldots R_{(n)}^{a_nb_n}(y_n, q_n)  G_{a_1 \ldots a_n, b_1\ldots b_n}(x_1 - \frac{1}{2}y_1, \ldots, x_n - \frac{1}{2}y_n; x_1 + \frac{1}{2}y_1, \ldots, x_n - \frac{1}{2}y_n). \hspace{0.2cm} \label{probdenN}
\eey
Here $R^{(i)}(x, q)$ is the measurement kernel for the $i$-th detector. The field correlation function on the probability density, $G_{a_1 \ldots a_n, b_1\ldots b_n}(x_1, \ldots, x_n; x_1', \ldots, x_n')$, is given by
\bey
G_{a_1 \ldots a_n, b_1\ldots b_n}(x_1, \ldots, x_n; x_1', \ldots, x_n') = \langle \psi| {\cal T}^*[\hat{C}_{b_1}^{(1)}(x'_1) \ldots \hat{C}_{b_n}^{(n)}(x'_n) ]
\nonumber \\
\times {\cal T} [\hat{C}_{a_n}^{(n)}(x_n) \ldots \hat{C}_{a_1}^{(1)}(x_1)]|\psi\rangle \label{correl}
\eey
where ${\cal T}^*$ stands for reverse time ordering.

Again we note that in the appropriate regime, the probability becomes independent of the switching functions, and equal to
\bey
P_n(x_1,q_1; x_2, q_2; \ldots; x_n, q_n) = \int d^4 y_1 \ldots d^4 y_n     R_{(1)}^{a_1b_1}(y_1, q_1) \ldots R_{(n)}^{a_n1b_n}(y_n, q_n)
\nonumber \\
\times G_{a_1 \ldots a_n, b_1\ldots b_n}(x_1 - \frac{1}{2}y_1, \ldots, x_n - \frac{1}{2}y_n; x_1 + \frac{1}{2}y_1, \ldots, x_n + \frac{1}{2}y_n) \label{probden4}
\eey
\subsection{The detector kernel}
The contribution of each detector to the probability is determined by the detector kernel $R^{ab}(x, q)$, defined by Eq. (\ref{detkern}). The detector kernel coincides with the matrix elements of a unitary operator,
\bey
R^{ab}(x, q) = \langle a, q|e^{i \hat{p} \cdot (x - x_0)}|b, q\rangle,
\eey
where $|a, q\rangle = \sqrt{\hat{\Pi}}(q) \hat{J}^a(x_0)|\Omega\rangle$.  If $\hat{\Pi}$ is a projector, then vectors $|a, q\rangle$ with different values of $q$ are orthogonal: $\langle a, q| b, q'\rangle = 0$ for $q \neq q'$.

The Fourier transform of the detector kernel
\bey
\tilde{R}^{ab}(\xi, q) = \int d^4x e^{-i \xi\cdot x} R^{ab}(x, q),
\eey
is given by
\bey
\tilde{R}^{ab}(\xi, q) = (2\pi)^4 e^{i \xi\cdot x_0} \langle a, q|\hat{E}_{\xi}|b, q\rangle,
\eey
where $\hat{E}_{\xi} = \delta^4(\hat{p} - \xi)$ is the projector onto the subspace with four-momentum $\xi^{\mu}$,  $\langle a, q|\hat{E}_{\xi}|a, q\rangle \geq 0$.
The momentum four vector associated to the detector is timelike and the associated energy $p^0$ is positive. This means that $\tilde{R}^{ab}(p, q) = 0$ for spacelike $p$, or if $p^0 < 0$.

For a  scalar composite operator $\hat{C}(x)$, we can drop the indices $a, b, \ldots$ from the detector kernel and write simply $R(x, q)$. If we record no other observable, we simply write $R(x) := \langle \omega|e^{i \hat{p} \cdot (x - x_0)}|\omega\rangle$, where $|\omega\rangle =  \hat{J}^a(x_0)|\Omega\rangle$. Hence, the detector kernel is determined by the energy-momentum distribution of the `threshold' state $\omega$.

\section{The collapsing shell model in 1+1-dimensional spacetime }

We consider the model of a collapsing null shell in a 1+1-dimensional spacetime \cite{Unruh76, BiDa}. The metric is:
\\
\bey
\left\{
\begin{array}{ll}
ds^2_{in} = A(\tau,r ) \Big( d\tau^2 - dr^2\Big),   \\
ds^2_{out}=C(r) \; dt^2 - C^{-1}(r)\; dr^2 =  C(r)\Big( d\tau^2 - dr^{*2}\Big). \label{metric}
\end{array}
\right.,
\eey
where $r^*$ is the tortoise coordinate, defined by $ dr^* = C^{-1}(r)\; dr$.

Without loss of generality, we assume $A(t, r) = 1$  \cite{Dav}.  We take $C(r)=(1-\dfrac{2M}{r})$, where $M$ is the mass of the shell, hence  $C \rightarrow 1 $ and $C' \rightarrow 0$ as $r \rightarrow \infty$. The horizon at $r=r_h$ corresponds to $C(r_h)=0$.    

Using the interior coordinate $\tau$ as a time parameter, we express the radius of the collapsing shell as $R(\tau)$. We assume that the collapse begins at $\tau = 0$, so $R(\tau) = R_0 > 2M$ for $\tau < 0$. The simplest form of collapse is linear: $ R(\tau)=R_0 - \nu \tau$. Then, the horizon forms at the proper time  $\tau_h= (R_0 -2M)/\nu$.

We  connect the timelike coordinates for the interior and exterior solution by imposing the continuity relation for the metric at $r=R$, 
\begin{align}\label{time coords}
\left(\dfrac{dt}{d\tau}\right)^2 = C(R)^{-1} -\left(C(R)^{-1}-C(R)^{-2}\right) \left(\dfrac{dR}{d\tau}\right)^2
\end{align}

 We introduce the incoming ($v,V$) and outgoing ($u,U$) null coordinates
\begin{equation}
\left\{
\begin{array}{ll}
U = \tau-r+R_0 \;\;, \;\; V= \tau + r -R_0  \;\;\;\;(inside) \\\\
u = t-r^*+R^*_0 \;\;, \;\; v= t + r^* -R^*_0\;\;\;(outside)
\end{array}
\right.
\end{equation}
defined so that  $u = U = v= V= 0$ at the surface of the shell as the collapse begins. Then, $ds^2_{in} =    dU \,dV $ and $ds^2_{out}=C(r) \; du \,dv$.

Next we consider   a massless scalar field obeying the Klein-Gordon equation $\;\square\phi=0$. Equivalently,  $\partial^2_{UV}\phi  = 0 $ inside the shell, and  $\partial^2_{uv}\phi = 0 $ outside the shell. The solutions to the field equations are
\bey
\phi = \left\{\begin{array}{ll}
 F(V) + G(U) \;,\; r\leq R(\tau)\;\;\;\;(inside)\\[2mm]
 f(v)+g(u)  \;\;\;\;,\; r\geq R(\tau)\;\;\;\;(outside)
\end{array}
\right.
\eey
for arbitrary functions $F,f$ for the incoming   and $G,g$ for the outgoing component. For simplicity, we assume Dirichlet boundary conditions at $r = 0$, which imply that
\begin{equation}\label{Dirichlet phi}
  G(U)=-F(U-2R_0)
\end{equation}
We express the relation between the interior and exterior coordinates by the functions $U(u)$ and $v(V)$. Field continuity across the shell implies that 
 $F(V) = f\bigl(v(V)\bigr)$ and $g(u) = G\bigl(U(u)\bigr)$. Using Eq.  \eqref{Dirichlet phi} we find that 
\begin{equation}
g(u) = G\bigl(U(u)\bigr)
     = -F\bigl(U(u)-2R_0\bigr)
     = -f\Bigl(v\bigl(U(u)-2R_0\bigr)\Bigr).
\label{eq:g-from-f}
\end{equation}
Hence, the exterior field is
\begin{equation}
\phi(u,v) 
                       = f(v) - f\Bigl(v\bigl(U(u)-2R_0\bigr)\Bigr).
\label{fasas}
\end{equation}



%

To evaluate the functions $U(u)$ and $v(V)$, we match the coordinate systems along the shell's worldline. At the 
  limit of a near–null collapse, $\nu\to1$, we obtain $dv \simeq dV$, leading to $v(V) = V$, and 
\bey
 \frac{dU}{du} = \frac{U - U_h}{U - U_h -4M}
\eey
  where $U_h\equiv U(\tau_h)= 2R_0-4M$ is the value of $U$ at horizon formation.

The solution is
\begin{equation}
U(u) = U_h - 4M\,W_0\!\left(
  \frac{U_h}{4M}
  e^{\frac{U_h}{4M}}e^{-\frac{u}{4M}}
\right),
\label{eq:U-of-u}
\end{equation}
where $W_0$ is the Lambert function---the inverse of the function $f(x) = xe^x$---evaluated at its principal branch.

At the limit $u\to \infty$,   $U(u) \simeq U_h - A\,e^{-u/4M}$ for some constant $A$, and $U$ becomes essentially identical with the Kruskal coordinate in eternal black holes.
 
The solutions (\ref{fasas}) become 
  $\phi (u,v) = f(v)- f(p(u))$, where  
\begin{equation}\label{pu}
    p(u)= U(u)-2R_0 = -4M\left( W_0[D_0e^{-\frac{u}{4M}}] +1 \right)\;\;,D_0=\dfrac{U_h}{4M}e^{\frac{U_h}{4M}}
\end{equation}

\section{Derivation of the Wightman function}

To quantize the field, we choose as incoming modes the usual set of positive-frequency solutions that are purely incoming at past null infinity $\mathscr I^{-}$ and define the `\textit{in}' vacuum state as $\hat{a}_\omega \; \ket{0_{in}} = 0$, for all $\omega$. These are the normal modes that depend only on the advanced time $v$ as:
\begin{equation}
f_\omega(v) = \frac{1}{\sqrt{4\pi\omega}}\,e^{-i\omega v}, 
\qquad \omega>0,
\label{eq:in-modes}
\end{equation}
So the mode solutions are:
\begin{equation}\label{modes}
\phi_\omega (u,v)=\frac{1}{\sqrt{4\pi \omega}} \left( \;e^{-i\omega v} - e^{-i\omega p(u)} \right) 
\end{equation}
The field operator is then expanded in this complete set of modes as :
\begin{equation}\label{field quant}
    \hat\phi(u,v)=\sum_\omega \left( \hat{a}_\omega\; \phi_\omega(u,v)+\hat{a}_\omega^{\dagger} \; \phi_\omega^*(u,v) \right)
\end{equation}
with $\hat{a},\hat{a}^{\dagger}$ the standard annihilation and creation operators, obeying the usual commutation relations.

 The two-point correlation (Wightman) function in the `\textit{in}' vacuum is
\begin{equation}
  G^{(1,1)}(u,v;u',v')
  \equiv \langle 0_{\text{in}} | \phi(u,v)\,\phi(u',v') | 0_{\text{in}} \rangle.
\end{equation}
noting that  $a_{\omega}|0_{\text{in}}\rangle = 0$ and  \(
  \langle 0_{\text{in}} | a_{\omega} a_{\omega'}^{\dagger}
  | 0_{\text{in}} \rangle = \delta_{\omega \omega'}
\), we obtain
\begin{align*}
    G^{(1,1)}(u,v;u',v') &=  \sum_\omega\left( \phi_\omega(u,v) \; \phi_{\omega}^*(u',v') \right) 
    \nonumber \\
    &= \sum_\omega \dfrac{1}{4 \pi \omega}\left( e^{-i\omega(v-v')} + e^{-i\omega\left(p(u)-p(u')\right)} -  e^{-i\omega\left(p(u)-v'\right)} - e^{-i\omega\left(v-p(u')\right)} \right).
\end{align*}
 
In the continuous limit, we can replace the summation with integration and the two-point function is expressed in the integral form:
\begin{equation}
    G^{(1,1)}(u,v;u',v') =\int_0^{\infty} \dfrac{ d\omega}{ 4 \pi \omega}\left( e^{-i\omega(v-v')} + e^{-i\omega\left(p(u)-p(u')\right)} -  e^{-i\omega\left(p(u)-v'\right)} - e^{-i\omega\left(v-p(u')\right)} \right) .
\end{equation}
which yields
\begin{equation}\label{two-point function}
     G^{(1,1)}(u,v;u',v') =-\dfrac{1}{4 \pi} \ln\left(\dfrac{(v-v'-i\varepsilon)\left( p(u)-p(u')-i\varepsilon\right)}{(p(u)-v'-i\varepsilon)\left( v-p(u')-i\varepsilon\right)} \right)
\end{equation}
\\with the  $i\varepsilon$ prescription ensuring that the two-point function has the correct analytic structure for a Wightman function and regulates any singularities.

\section{Derivation of the single-time detection probabilities}
The QTP probability density $P_1(t, E)$ for a single detection event, recorded by a pointlike detector following a spacetime trajectory $x(t)$, is given by:
\begin{equation}\label{P1}
    P_1(t, E) = \int ds e^{-iEs} g(s) G^{(1,1)}(x(t + \frac{s}{2}); x(t - \frac{s}{2}))
\end{equation} 
By Eq.\eqref{two-point function}, we get
\begin{align}
    G^{(1,1)}(X(t + \frac{s}{2}); X(t - \frac{s}{2}))
    &=-\dfrac{1}{4 \pi}\Big[\ln(s-i\varepsilon)
    +\ln\big(p(t+\dfrac{s}{2}) -p(t-\dfrac{s}{2})-i\varepsilon\big) \nonumber \\
    &-\ln(t+\dfrac{s}{2} + r_0^*-R_0^* -p(t-\dfrac{s}{2})-i\varepsilon)  
    -\ln( p(t+\dfrac{s}{2}) - t +\dfrac{s}{2}-r_0^*+R_0^*-i\varepsilon)\Big]\nonumber
\end{align}
Substituting to \eqref{P1} yields four logarithmic integrals. Denoting the resulting integrals by their respective dependencies on the coordinates  $v,v',\tilde{U},\tilde{U}'$, where $\tilde{U}:=p(u)$, we get:
\begin{equation}
  P_1(t,E) = -\dfrac{1}{4 \pi} 
  \left( I^{(\Delta v)} + I^{(\Delta \tilde{U})} - I^{(\tilde{U} v)} - I^{(v\tilde{U})}\right)
\end{equation}
At the limit $E\sigma \gg 1$, where the effects of the switching function can be ignored
\begin{align}
     I^{(\Delta v)} &= \int_{-\infty}^{\infty} ds e^{-iEs}\;\ln(s-i\varepsilon)\\
     I^{(\Delta \tilde{U})} &= \int_{-\infty}^{\infty} ds e^{-iEs}\;\ln\big(p(t+\dfrac{s}{2}) -p(t-\dfrac{s}{2})-i\varepsilon\big)\\
      I^{(\tilde{U} v)} &= \int_{-\infty}^{\infty} ds e^{-iEs}\;\ln(t+\dfrac{s}{2} + r_0^*-R_0^* -p(t-\dfrac{s}{2})-i\varepsilon)\\
       I^{(v \tilde{U} )} &= \int_{-\infty}^{\infty} ds e^{-iEs}\;\ln( p(t+\dfrac{s}{2}) - t +\dfrac{s}{2}-r_0^*+R_0^*-i\varepsilon)
\end{align}

\begin{figure}[h!]
    \centering
    \includegraphics[width=0.6\textwidth]{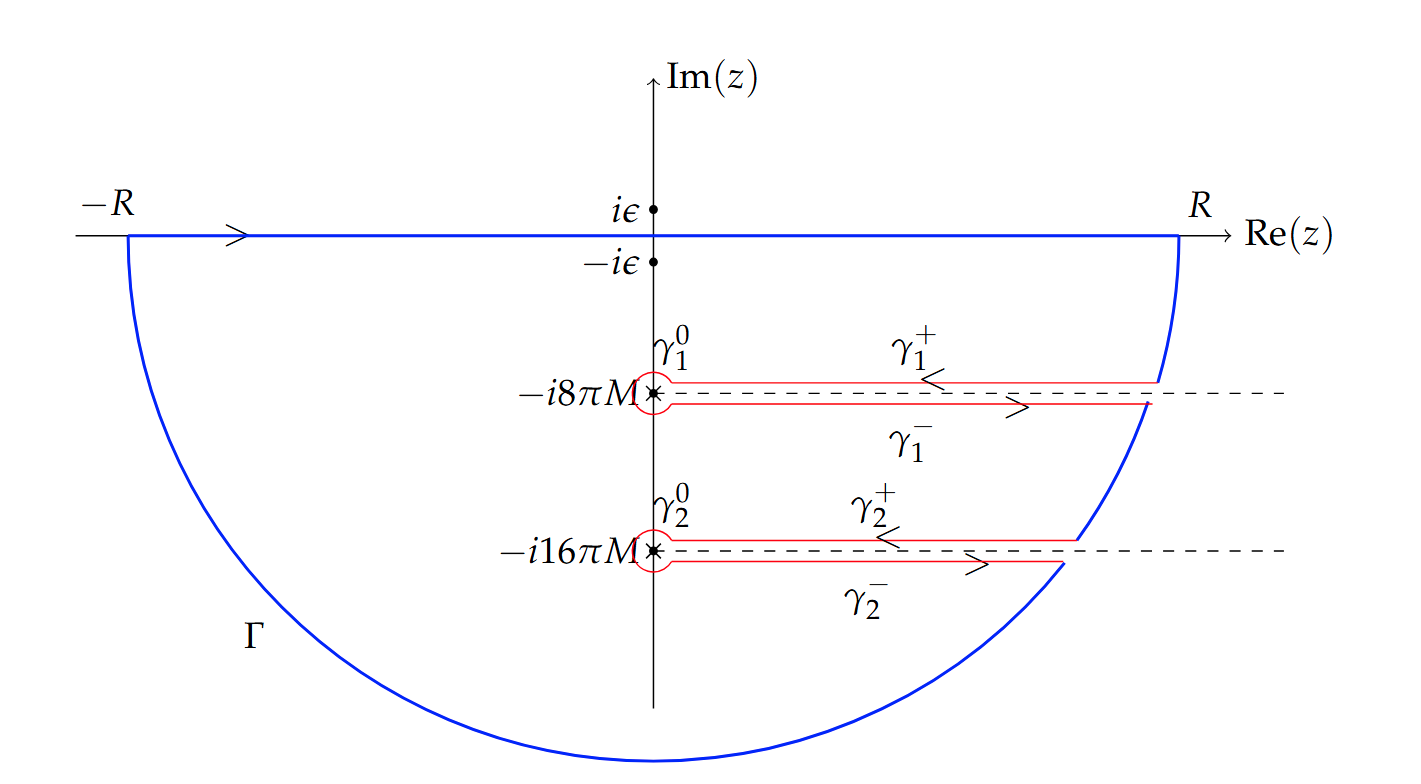}
    \caption{The contour used for the integration of $I^{(\Delta \tilde{U})}.$ We consider the lower half plane where we avoid the branch cuts created by the branch points $z_k=i(8\pi M)k$.}
    \label{I2Contour}
\end{figure}

The first term $I_1$ is related to the Minkowski Wightman term for an inertial detector and as such its contribution is zero. The integrals  $I^{(\tilde{U} v)}$ and $I^{(v \tilde{U} )}$  also vanish as a consquence of the integral identity
\bey
\int_{-\infty}^{\infty} ds\; e^{-iEs}\,\ln \big(F(s)-i\varepsilon\big) = \begin{cases} 
0, &   F'(s_0) > 0 \\ 
-\frac{2\pi}{E} e^{-iEs_0}, &   F'(s_0) < 0.
\end{cases} 
\eey
for a function $F(s)$ with only one real, simple root $s_0$, and $E>0$
. The only non-vanishing term is $I^{(\Delta \tilde{U})}$, and this term yields the Planck thermal spectrum:
\begin{equation}
    P_1(t,E)=\dfrac{1}{2E\left( e^{8\pi ME}-1\right)} \label{pite}
\end{equation}
To show this, 
we rewrite $I^{(\Delta \tilde{U})}$, as

\begin{equation}\label{IDU}
    I^{(\Delta \tilde{U})}=\int_{-\infty}^{\infty} ds e^{-iEs}\; \ln\bigg(-4M\big(\mathcal{W}\big[ \bar{D_0}e^{-\frac{s}{8M}} \big]-\mathcal{W}\big[ \bar{D_0}e^{\frac{s}{8M}} \big] \big)-i\varepsilon\bigg),
\end{equation}
The integrand has branch-cuts at $z = z_k = i(8\pi M)k\;,\;k \in \mathbf{Z}$. We  choose a contour $C$ that avoid  the branch points and the corresponding branch cuts, as in Fig.\ref{I2Contour}. We also  exclude the $z=0$ contribution due to the term $-i\varepsilon$ that shifts the root into the upper half plane. Then, using  Cauchy's theorem, we obtain
\begin{equation}
 I^{(\Delta \tilde{U})} + \frac{2\pi}{E} \sum_{k=1}^{\infty} \mathrm{e}^{-8\pi MEk} = 0. \label{eq:95}
\end{equation}
The   geometric series converges to $(\mathrm{e}^{8\pi ME} - 1)^{-1}$, and Eq. (\ref{pite}) follows.

\bigskip

\section{Derivation of the two-time detection probabilities}
The core calculation in this work is the evaluation of the joint probability density $P_2(t_1, E_1; t_2, E_2) $ of detection given by Eq. (3) in the main text. This depends on the four-point function 
$G^{(2,2)}(x_1, x_2; x'_1, x'_2)$---given by Eq. (1) of the main text for $\hat{O}(x) = \hat{\phi}(x)$.

\begin{figure}[h]
    \centering
    \includegraphics[width=0.45\textwidth]{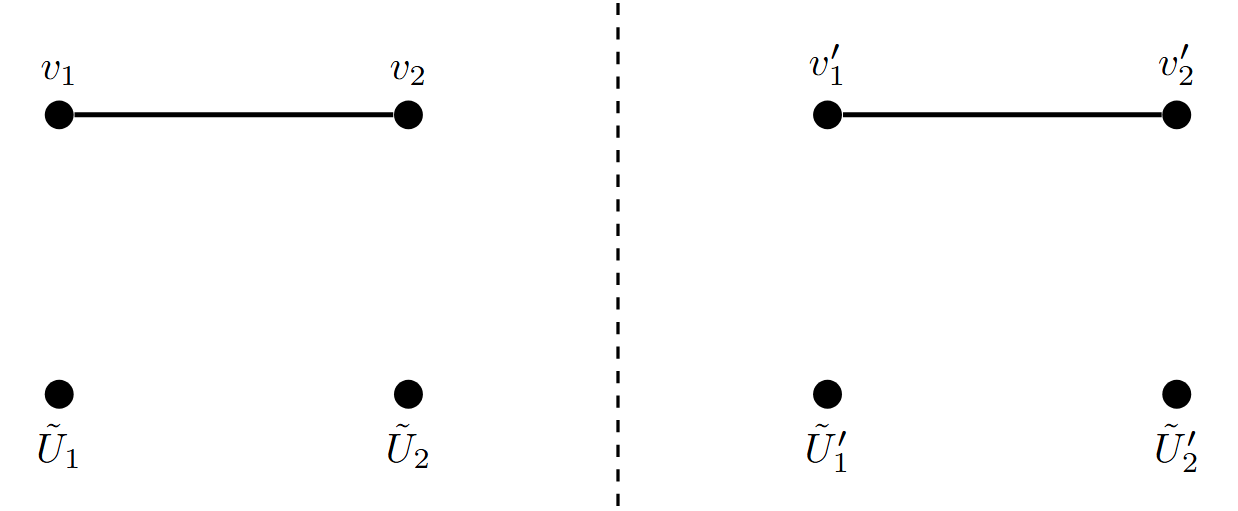}
    \caption{Graphical representation of the integral $I_A^{(\Delta v)(\Delta v)}, $ corresponding to a combination of the pairs $(v_1,v_2)$ and $(v_1',v_2')$. }
    \label{DotsDvDv}
\end{figure}

By Wick's theorem,
\begin{equation*}
    G^{(2,2)}(x_1, x_2; x_1', x_2') = G^{(1,1)}(x_1, x_1') G^{(1,1)}(x_2, x_2') + 
G^{(2,0)}(x_1, x_2) G^{(0,2)}(x_1', x_2') + G^{(1,1)}(x_1, x_2') G^{(1,1)}(x_2, x_1')
\end{equation*}
The first term generates the product $P_1(t_1, E_1) P_1(t_2, E_2)$. 
The second and the third term generate what we call the the A- and B-channel contributions to the joint probability density, 
\begin{align*}
    P_{2}^{A(B)}(t_1, E_1; t_2, E_2) =  \int_{-\infty}^{\infty}ds_1\int_{-\infty}^{\infty}ds_2\;\mathcal{K}(s_1,s_2)\;G_{A(B)}^{(2,2)}(x(t_1+\frac{s_1}{2}),x(t_2+\frac{s_2}{2});x(t_1-\frac{s_1}{2}),x(t_2-\frac{s_2}{2})),
    \end{align*}
where  $\mathcal{K}(s_1,s_2):=e^{-iE_1s_1} e^{-iE_2s_2} g(s_1)g(s_2)$ and 

\begin{equation}\label{ABchannels}
\left\{
\begin{array}{ll}
G_A^{(2,2)}:=  \frac{1}{16\pi^2}\ln \left( \frac{(v_1 - v_2 - i\varepsilon) (p(u_1) - p(u_2) - i\varepsilon)}{(v_1 - p(u_2) - i\varepsilon)(p(u_1) - v_2 - i\varepsilon)} \right) \ln \left( \frac{(v_1 ' - v_2 ' - i\varepsilon) (p(u_1 ') - p(u_2 ') - i\varepsilon)}{(v_1 ' - p(u_2 ') - i\varepsilon)(p(u_1 ') - v_2 ' - i\varepsilon)} \right)  \\\\
G_B^{(2,2)}:=  \frac{1}{16\pi^2}\ln \left( \frac{(v_1 - v_2' - i\varepsilon) (p(u_1) - p(u_2') - i\varepsilon)}{(v_1 - p(u_2') - i\varepsilon)(p(u_1) - v_2' - i\varepsilon)} \right)
\ln \left( \frac{(v_2 - v_1 '  - i\varepsilon) (p(u_2 ) - p(u_1 ')  - i\varepsilon)}{(v_2 - p(u_1 ') - i\varepsilon)(p(u_2)  - v_1 ' - i\varepsilon)} \right) 
\end{array}
\right.
\end{equation}
 Each channel  requires the evaluation of sixteen integrals, each integral involving the product of two logarithms. We can express these integrals diagrammatically by noting that the four-point function $G^{(2,2)}$ involves eight coordinates: $v_1, v_2, v_1', v_2', \tilde{U}_1, \tilde{U}_2, \tilde{U}'_1,$ and $\tilde{U}'_2$, where $\tilde{U}=p(u)$. Each integral to be evaluated involves one pair of such coordinates in each of the two logarithms. Hence, it can be represented graphically by two lines connecting the pair of coordinates in each logarithm. For example, the integral 
 
\begin{align*}
    I_A^{(\Delta v)(\Delta v)} &:= \int_{-\infty}^{\infty}ds_1\int_{-\infty}^{\infty}ds_2\;\mathcal{K}(s_1,s_2) 
\ln \left(v_1-v_2 - i\epsilon \right)
\ln \left(v_1'-v_2'- i\epsilon \right) 
\end{align*}
is represented by the diagram of Fig. \ref{DotsDvDv}.

\subsection{Unruh-vacuum terms}
\begin{figure}[h]
    \centering
    \includegraphics[width=0.475\textwidth]{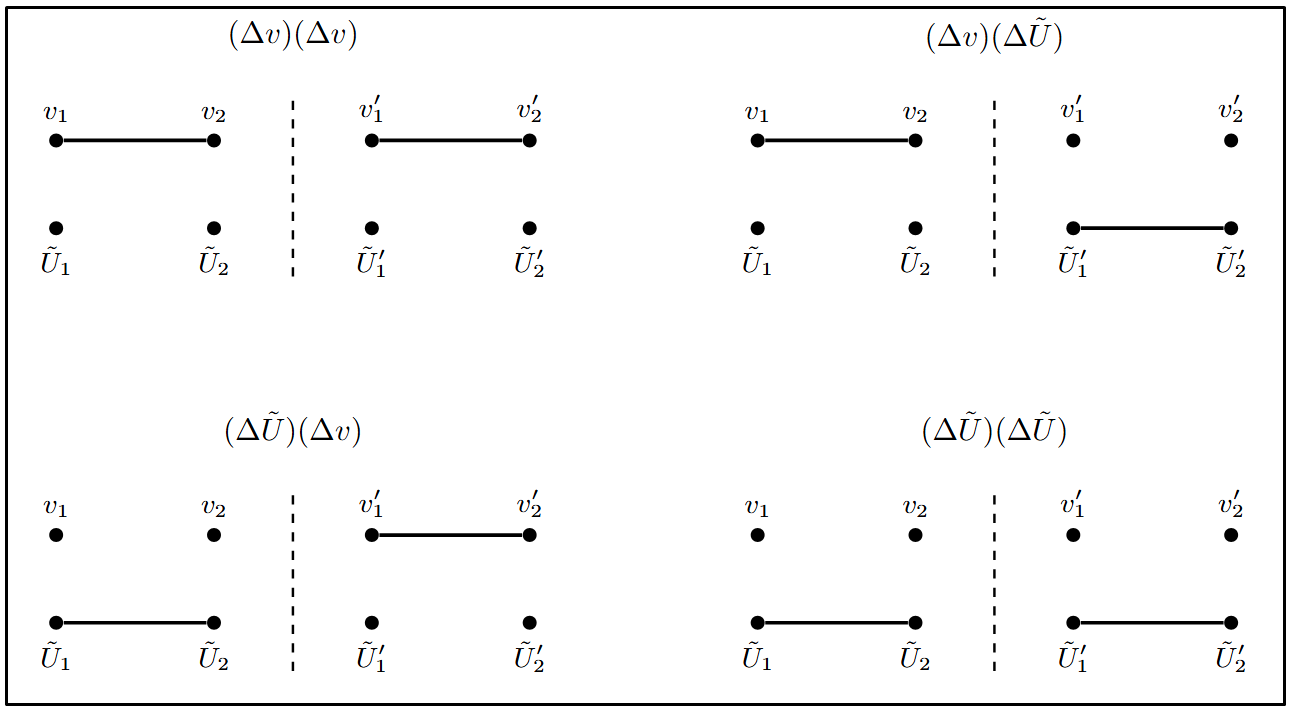}
     \includegraphics[width=0.49\textwidth]{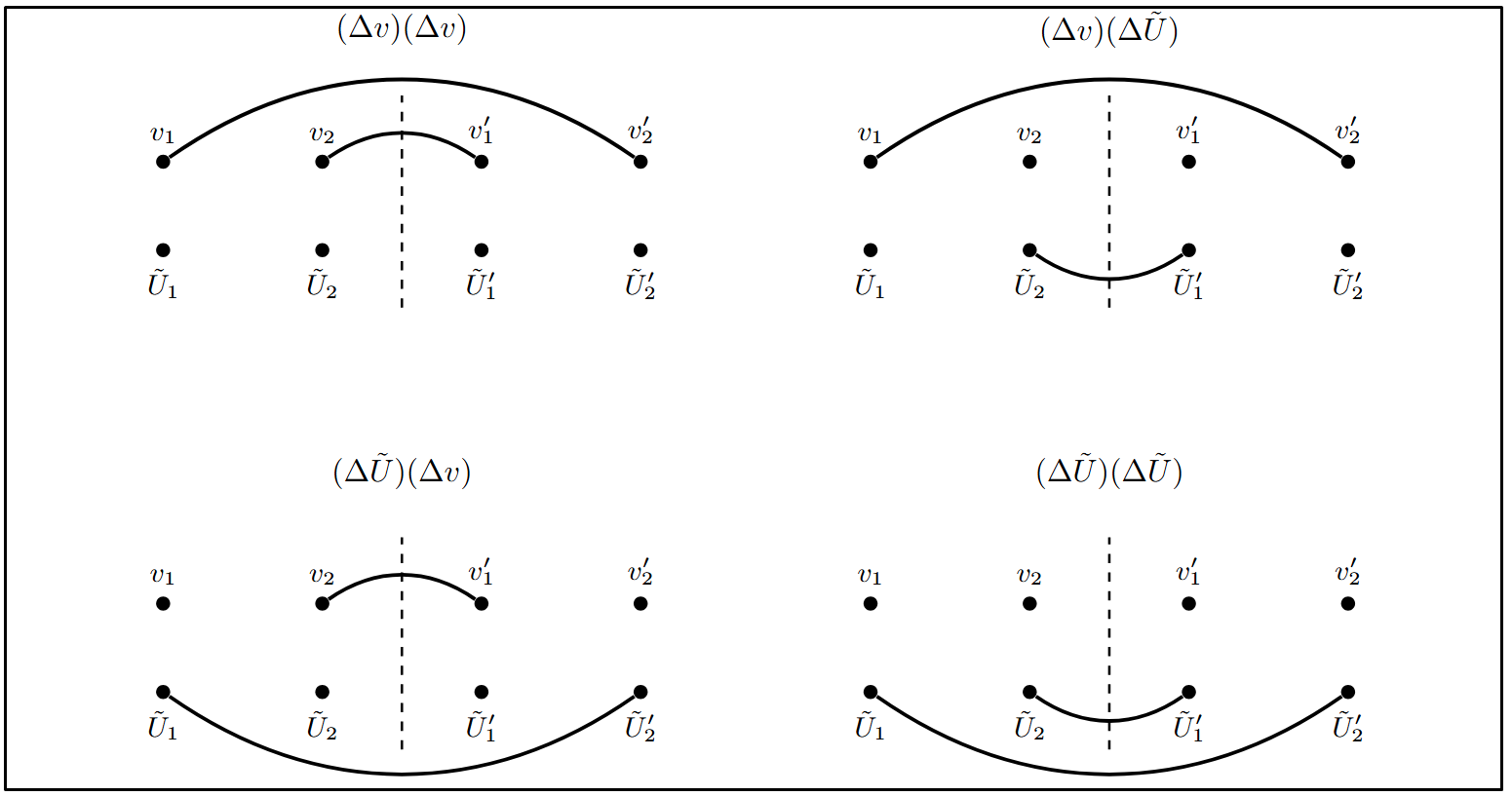}
    \caption{ The   Unruh-vacuum terms   of the A-channel (left) and of the    B-channel (right).}
    \label{DotsUnruh}
\end{figure}

Some terms in the joint probability originate from the two-point function of the  Unruh vacuum
\begin{equation}
    G_U^{(2)}(x, x') \propto \ln[(\Delta U - i\epsilon)(\Delta v - i\epsilon)],
\end{equation}
where $U:=-\kappa^{-1}\,e^{-\kappa u}$ is the Kruskal coordinate associated with the retarded coordinate $u$ and $\kappa:=(4M)^{-1}$, the surface gravity of the horizon. Hence, any integral in the probability density $P_2(t_1,E_1;t_2,E_2)$ containing both $\Delta \tilde{U}$ and $\Delta v$ terms corresponds to the Unruh vacuum. There are eight such terms, whose diagrams are shown in Fig. \ref{DotsUnruh}.

It is straightforward to show that the $A-$channel terms are supported on $E_1+E_2=0$ whereas the $B$-channel ones are supported on $E_1-E_2=0$. At the limit  $E \sigma \gg1$, these translate to the Dirac delta functions $\delta(E_1+E_2)$ and $\delta(E_1-E_2)$ correspondingly. Since $E_1,E_2>0$ we only need to focus on the $B-$channel terms.

\begin{figure}[t!]
\includegraphics[width=0.45\textwidth]{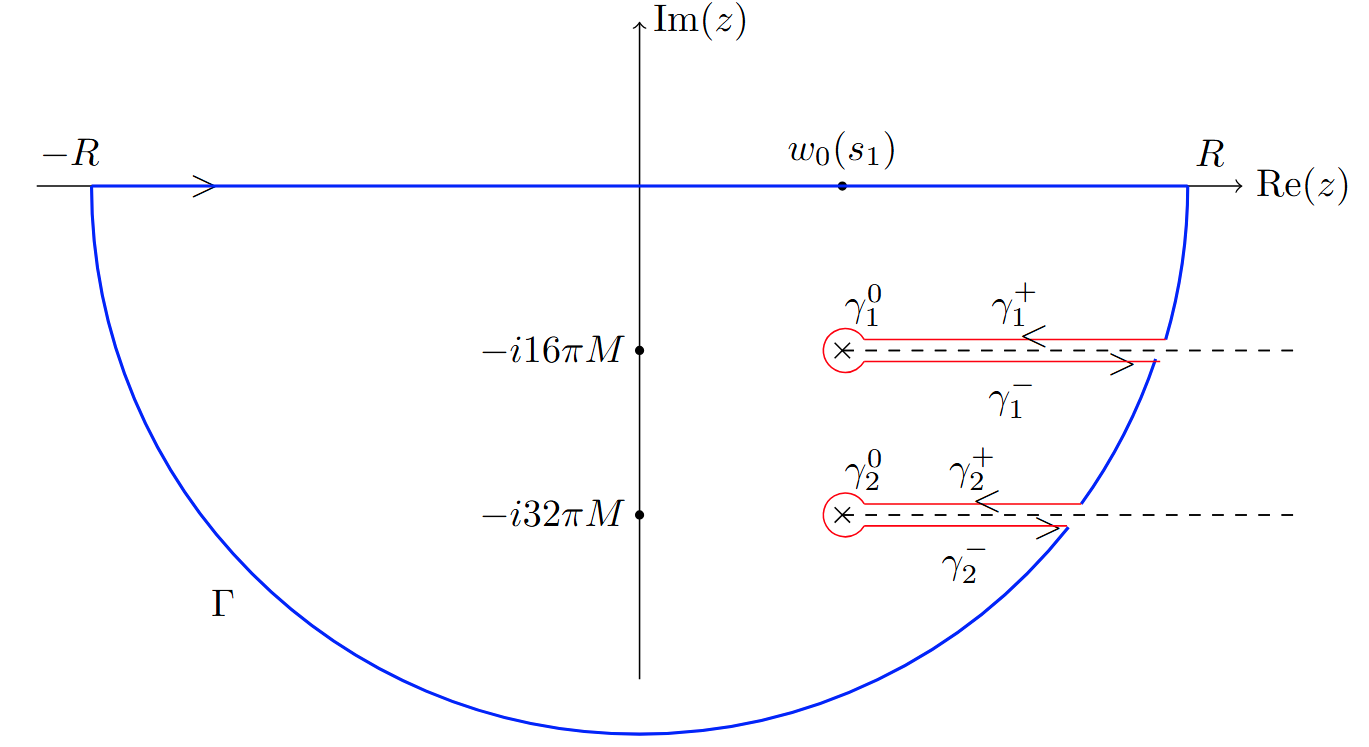}
    \includegraphics[width=0.465\textwidth]{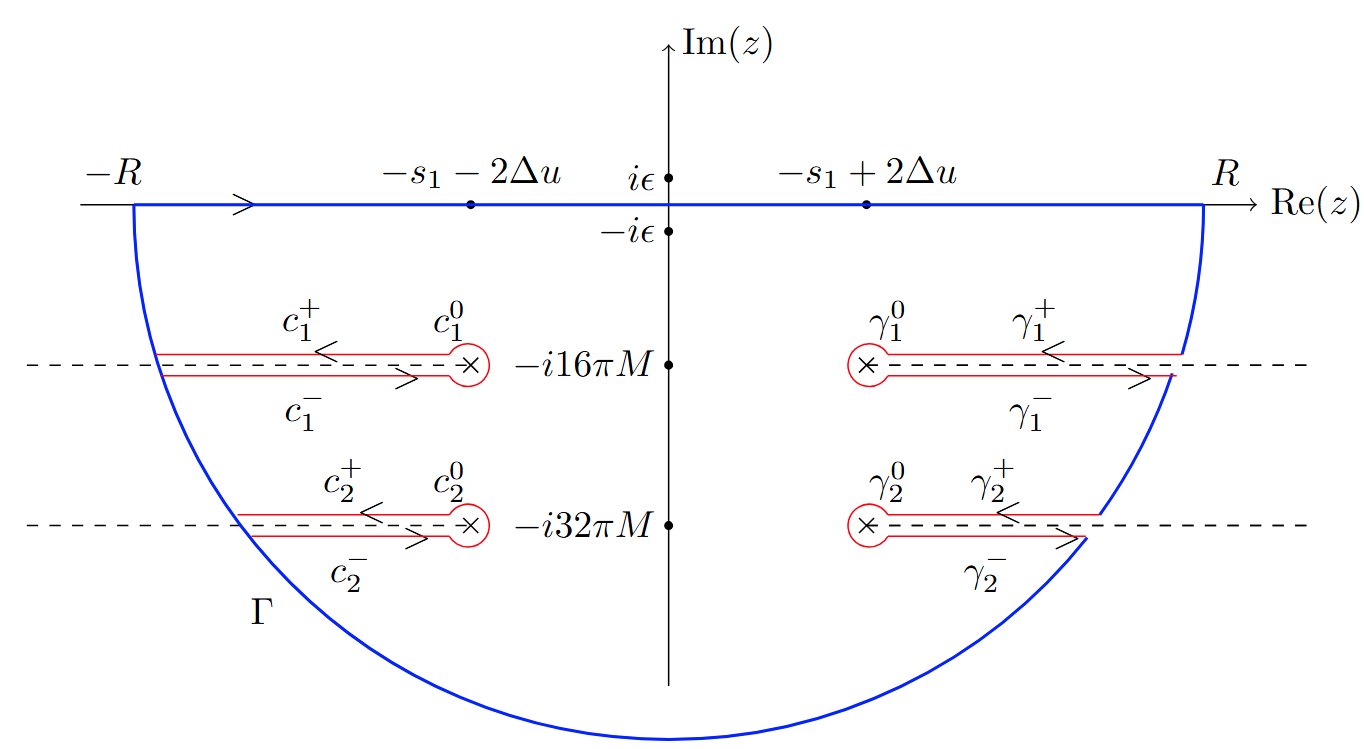}
\caption{\textbf{Left:} The closed contour used for the complex contour integration of the conjugate integrals $I_A^{(\Delta \tilde{U})(\Delta v)}$ and $I_A^{(\Delta v)(\Delta \tilde{U})}$. The contour is closed in the lower half-plane by a large semi-circle $\Gamma$ of radius $R \to \infty$. The contribution of the large semicircle $\Gamma$ vanishes, due to Jordan's Lemma. We have deformed the contour in a way that avoids the branch points $w_0-i16\pi Mk$ for $k \in \mathbb{Z}$, as well as the branch cuts generated by those. This deformation consists of paths $\gamma_k^+$ and $\gamma_k^-$, which run parallel to the branch cuts, yielding the $\Delta J_{A_k}^{(\Delta \tilde{U})(\Delta v)}$ terms, and small circular arcs $\gamma_k^0$ that enclose the branch points, whose contributions vanish in the limit of their radii going to zero. \textbf{Right:} The closed contour used to evaluate the integral $ J_B^{(\Delta U)(\Delta U)}$. It follows the same logic, with the difference that the extra $(\Delta U)$ term on the logarithmic argument of the integrand yields a second family of branch points. The right-side branch cuts, starting at $(-s_1+2\Delta u)-i16\pi Mk$, are encircled by the paths $\gamma_k^\pm$ and the small arcs $\gamma_k^0$, while the mirrored left-side branch cuts, starting at $(-s_1-2\Delta u)-i16\pi Mk$, are treated analogously by the paths $c_k^\pm$ and the small arcs $c_k^0$.}
    \label{ResUnruh}
\end{figure}

The $(\Delta v)(\Delta v)$ term corresponds to the detector response in   the Minkowski vacuum and thus yields zero. The cross terms $(\Delta v)(\Delta U)$, $(\Delta U)(\Delta v)$ are symmetric under the swap $(E_1,E_2)\to(-E_1,-E_2)$, so they are complex conjugates under the same $i\varepsilon$ regulator convention. For $E \sigma \gg1$
\begin{align*}
    I_A^{(\Delta v)(\Delta \tilde{U})} &= \int_{-\infty}^{\infty}ds_1\int_{-\infty}^{\infty}ds_2\;e^{-i(E_1s_1+E_2s_2)}\ln \left( \frac{s_1-s_2}{2} - \Delta v - i\epsilon \right) \times\\
    &\qquad\qquad\qquad\qquad\times
\ln \left( -4 M \left( \mathcal{W} [D_1 e^{\frac{s_1}{8M}}] - \mathcal{W}[D_2 e^{\frac{s_2}{8M}}] \right) - i\epsilon \right) \\
     I_A^{(\Delta \tilde{U})(\Delta v)}&=\int_{-\infty}^{\infty}ds_1\int_{-\infty}^{\infty}ds_2\;e^{-i(E_1s_1+E_2s_2)}
\ln \left( -4 M \left( \mathcal{W} [D_1 e^{-\frac{s_1}{8M}}] - \mathcal{W}[D_2 e^{-\frac{s_2}{8M}}] \right) - i\epsilon \right)\times\\
&\qquad\qquad\qquad\qquad\times
\ln \left( -\frac{s_1-s_2}{2}-\Delta v - i\epsilon \right) 
\end{align*}
Integrals with at least one $\Delta U$ term  have integrands of the form  $ln(\mathcal{W}[\cdot]-\mathcal{W}[\cdot])$. These functions have infinitely many branch points of the form $w_k=w_0-i16\pi Mk$, for $k \in \mathbb{Z}$. In the terms above, the branch points are identified by  the roots of the equation:
\begin{equation}
    \mathcal{W} [D_1 e^{\frac{ s_1}{8M}}] - \mathcal{W}[D_2 e^{\frac{s_2}{8M}}]=0 \Rightarrow s_{2_,k}(s_1)= (s_1+2\Delta u)-i16\pi Mk,\;\; k \in \mathbb{Z}
\end{equation}
First, we integrate over $s_2$. We extend to the complex plane and conduct the contour integration: 
\begin{align*}
J_A^{(\Delta \tilde{U})(\Delta v)}(s_1) &= \oint_c dz \, e^{-iE_2z}  
\ln \left( \frac{s_1-z}{2} - \Delta v - i\epsilon \right)
\ln \left( -4 M \left( \mathcal{W} [D_1 e^{\frac{s_1}{8M}}] - \mathcal{W}[D_2 e^{\frac{z}{8M}}] \right) - i\epsilon \right),
\end{align*}
for the closed contour shown in Fig. \ref{ResUnruh}. Then, by Cauchy's  theorem, as the contour encloses no poles or branch points, we get:
\begin{align*}
  J_A^{(\Delta v)(\Delta \tilde{U})}(s_1) &= \int_{-\infty}^{\infty}ds_2\;e^{-iE_2s_2}
\ln \left( \frac{s_1-s_2}{2} - \Delta v - i\epsilon \right)
\ln \left( -4 M \left( \mathcal{W} [D_1 e^{\frac{s_1}{8M}}] - \mathcal{W}[D_2 e^{\frac{s_2}{8M}}] \right) - i\epsilon \right)\\
&+ \int_{\Gamma} dz  + \sum_{k=1}^\infty \left[ \left(\Delta J_{A_k}^{(\Delta v)(\Delta \tilde{U})} \right) + \int_{\gamma_k^0} dz \right]=0
\end{align*}
The integration over the large semicircle $\Gamma$ as well as the small circles $\gamma_k^0$ yields zero. We denote $\Delta J_{A_k}^{(\Delta v)(\Delta \tilde{U})}$ the contribution of the integral from the lines $\gamma_k^+$ and $\gamma_k^-$, i.e. the contribution of the integral slightly above and below the branch cut. Hence:
\begin{align*}
I_A^{(\Delta v)(\Delta \tilde{U})} :=\int_{-\infty}^{\infty} ds_1 e^{-iE_1 s_1}J_A^{(\Delta v)(\Delta \tilde{U})}(s_1)= \int_{-\infty}^{\infty} ds_1 e^{-iE_1 s_1}\left(- \sum_{k=1}^\infty\left(\Delta J_{A_k}^{(\Delta v)(\Delta \tilde{U})} \right)\right)
\end{align*}
The branch cut contribution terms are given by:
\begin{align*}
    \Delta J_{A_k}^{(\Delta v)(\Delta \tilde{U})}
    &=\int_{\gamma_k^+}ds_2\;e^{-iE_2s_2}\ln \left( \frac{s_1-s_2}{2} - \Delta v - i\epsilon \right)
\ln \left( -4 M \left( \mathcal{W} [D_1 e^{\frac{s_1}{8M}}] - \mathcal{W}[D_2 e^{\frac{s_2}{8M}}] \right) - i\epsilon \right)\Bigg|_{upper} \\
    &+\int_{\gamma_k^-}ds_2\;e^{-iE_2s_2}\ln \left( \frac{s_1-s_2}{2} - \Delta v - i\epsilon \right)
\ln \left( -4 M \left( \mathcal{W} [D_1 e^{\frac{s_1}{8M}}] - \mathcal{W}[D_2 e^{\frac{s_2}{8M}}] \right) - i\epsilon \right)\Bigg|_{lower}\\
    &=e^{-iE_2s_1}e^{-2iE_2\Delta u} e^{-(16\pi M  E_2)k} \cdot (2\pi i )  \int_{0}^{\infty} dx \,  e^{-iE_2 x} 
\ln \left( -\frac{x}{2} + 2(t_1 - t_2) + 8\pi M k i  \right) 
\end{align*}
The indices "upper" and "lower" indicate the branch of the logarithm depending on each relative position over the branch cut. Their difference yields a factor of $2\pi i$ over the remaining logarithmic integral. The remaining logarithmic integral is known, yielding a logarithmic term and an Exponential integral one. So, we get:
\begin{align}
   &I_A^{(\Delta \tilde{U})(\Delta v)} =\dfrac{2\pi}{E_2} e^{-2iE_2\Delta u}\int_{-\infty}^{\infty} ds_1 e^{-i(E_1+E_2) s_1}\left(- \sum_{k=1}^{\infty} e^{-(16\pi M  E_2)k} \ln(2(t_1 - t_2)+i8\pi M k) \right) \nonumber \\
   &\qquad+ \dfrac{2\pi}{E_2} e^{-2iE_2(\Delta u+2(t_1-t_2))}\int_{-\infty}^{\infty} ds_1 e^{-i(E_1+E_2) s_1}\left(- \sum_{k=1}^{\infty}\;\; \mathbf{E_1}[16\pi ME_2k-i4E_2(t_1-t_2)] \right)
\end{align}
We notice that the quantities inside the parentheses are independent of the integration variable. Hence, the integral reduces to $\delta(E_1+E_2)$ terms. Likewise, the B-channel analogues also yield Dirac delta $\delta(E_1-E_2)$ terms. Similar analysis gives:
\begin{align}
    I_B^{(\Delta \tilde{U})(\Delta v)}+I_B^{(\Delta v)(\Delta \tilde{U})}&=
    -\dfrac{(2\pi)^2}{E_1}\delta(E_1-E_2)\Big(Re\left\{e^{i2E_1\Delta u} S_1(16\pi ME_1;2\Delta t,8\pi M) \right\} \nonumber \\
    &\qquad\qquad+ Re\Big\{e^{i2E_1\Delta v}S_2(16\pi ME_1,4E_1\Delta t)\Big\} \Big)
\end{align}
where we have defined $ S_1(a;x,y):= \sum_{k=1}^\infty \ e^{-ak} \ln(x - iy k)$ and $S_2(x,y):= \sum_{k=1}^\infty\;\mathbf{E_1}[xk+iy] $

The $(\Delta \tilde{U})(\Delta \tilde{U})$ terms are Unruh-like, in the limit of $u\to +\infty$. So the corresponding integral from the Unruh vacuum four-point function is:
\begin{align}
J_B^{(\Delta U)(\Delta U)}
&=
\int 
\int ds_1 ds_2\,
e^{-iE_1 s_1-iE_2 s_2}
\ln\!\left(
-\left[C_1 e^{-s_1/(8M)}-C_2 e^{+s_2/(8M)}\right]-i\epsilon
\right)\times \nonumber\\
&\qquad\qquad\times
\ln\!\left(
-\left[C_2 e^{-s_2/(8M)}-C_1 e^{+s_1/(8M)}\right]-i\epsilon
\right).
\end{align}
Each of the logarithmic arguments produces infinitely many branch points located at $ p_k^{(\pm)}(s_1) = -s_1 \pm 2\Delta u - 16\pi i M k$, with $k \in \mathbb{Z}.$ So extending to complex contour integration, we deform the contour in a symmetric way as shown in the right part of Fig.\eqref{ResUnruh}. By Cauchy's theorem we get: 
\begin{equation*}
   J_B^{(\Delta U)(\Delta U)}=\int_{-\infty}^{+\infty}ds_1\,e^{-iE_1s_1}\Bigg(-\sum_{k=1}^\infty \Delta J_{c_k}-\sum_{m=1}^\infty \Delta J_{\gamma_m}\Bigg),
\end{equation*}
Ignoring the branch constants and the contact $\delta '(E_1-E_2)$ terms, it yields:
\begin{equation*}
   J_B^{(\Delta U)(\Delta U)} =-\frac{8\pi^2}{E_1(e^{16\pi M E_1} - 1)} \delta(E_1 - E_2) 
Re
\left\{
e^{-2iE_1|\Delta u|}\;
\mathbf{L}\left(e^{-|\Delta u|/(2M)}\,,\,i8ME_1\right)
\right\}.
\end{equation*}
where we have defined:
\begin{equation}
    \mathbf{L}(q,z):=
z\int_0^1 dt\,
t^{z-1}
\ln(1-qt-i\varepsilon),\;\;\varepsilon\to 0^+
\end{equation}
Thus, the second-order coherence function for the  Unruh-vacuum contribution takes the form:
\begin{align*}
     &C^{(2)}_U(t_1, E_1; t_2, E_2):= \frac{P_2^{(U)}(t_1, E_1; t_2, E_2) }{P_1(t_1, E_1) P_1(t_2, E_2)}= \frac{\frac{1}{16\pi^2}\left(I_B^{(\Delta v)(\Delta v)}+I_B^{(\Delta \tilde{U})(\Delta v)}+I_B^{(\Delta v)(\Delta \tilde{U})}+I_B^{(\Delta \tilde{U})(\Delta \tilde{U})}\right) }{P_1(t_1, E_1) P_1(t_2, E_2)}\\
     &=-\dfrac{E_1}{\mathbf{Z}^2[E_1]}\delta(E_1-E_2)\Bigg(Re\left\{e^{i2E_1\Delta u} S_1(16\pi ME_1;2\Delta t,8\pi M) \right\}+ Re\Big\{e^{i2E_1\Delta v}S_2(16\pi ME_1,4E_1\Delta t)\Big\}  \\
     &\qquad\qquad\qquad\qquad\qquad\qquad+ \mathbf{Z}[E]Re\left\{
e^{-2iE_1|\Delta u|}\;
\mathbf{L}\left(e^{-|\Delta u|/(2M)}\,,\,i8ME_1\right)
\right\}    \Bigg) 
\end{align*}
\begin{figure}[b!] 
    \includegraphics[width=0.3\textwidth]{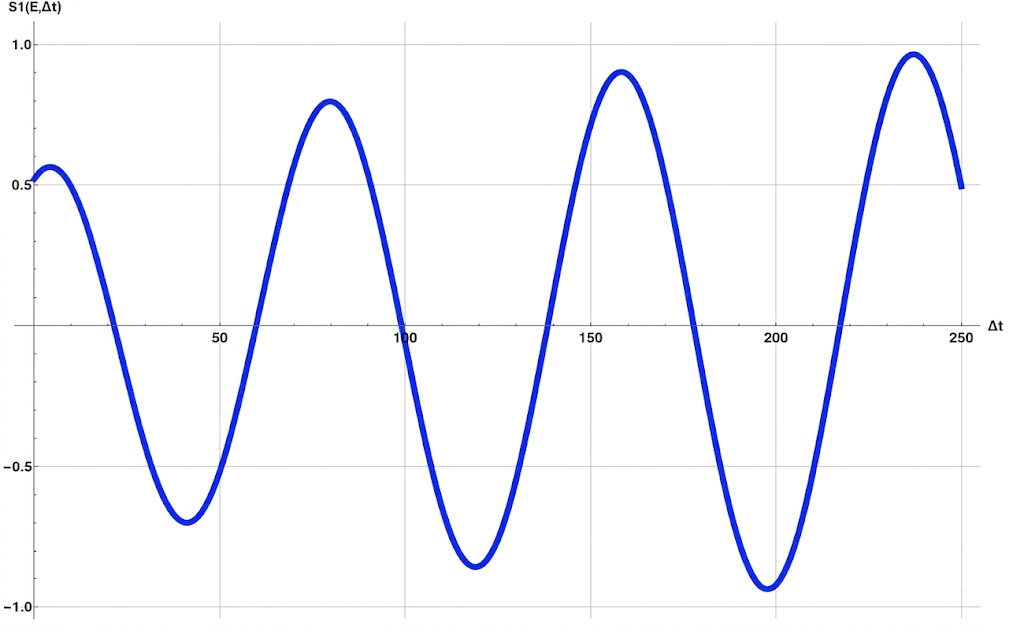}
     \includegraphics[width=0.3\textwidth]{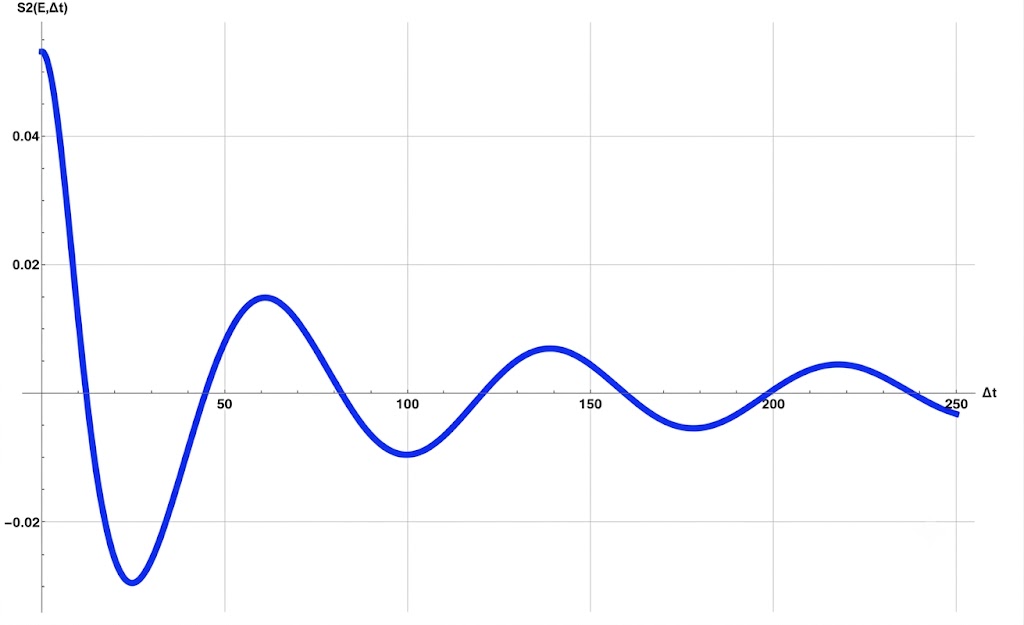}
     \includegraphics[width=0.3\textwidth]{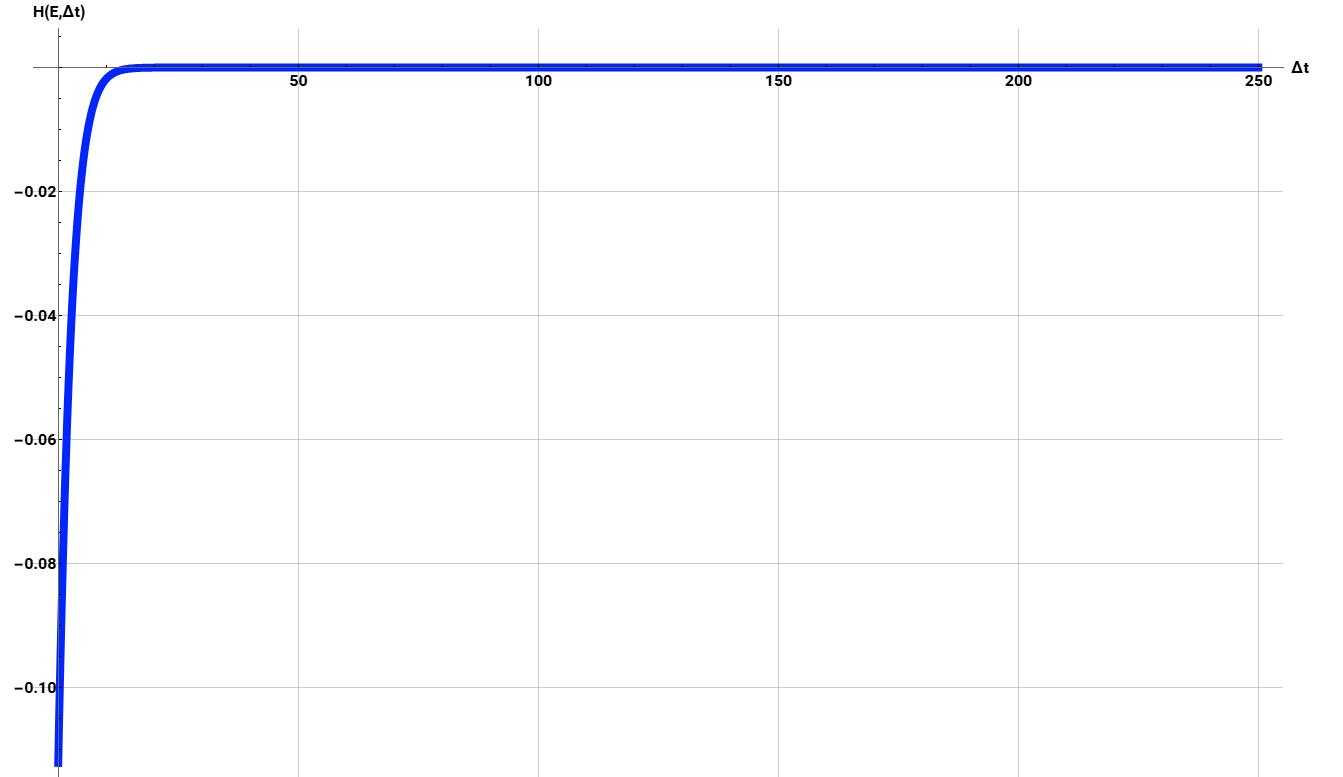}
\caption{Plot of the special functions $S_1$, $S_2$ and $\mathbf{L}$ for $E_1=T_h:=1/8\pi M$. We consider each detector at a fixed positions $r_1^*=r_2^*$, so they effectively become functions of $E_1$ and $\Delta t$, appearing in the same order. The plots run for $\Delta t \in (0,250)$, but they are symmetric over the $y-$axis, meaning for $\Delta t \to-\Delta t$, retaining the symmetry for the index swap $1 \leftrightarrow$ 2.}
    \label{Sum2Plot}
\end{figure}

\subsection{Non-Unruh terms}

For the remaining integrals, recognize two cases; the ones that contain a $\Delta \tilde{U}$ term, and as such yield contributions from the infinitely many branch points, and the ones that don't contain such terms, which are proven to be transient.

For the first case, in a similar manner to the Unruh-like terms, the integrals split to a logarithmic contribution and an Exponential-Integral one. The $A$-channel terms are identically zero for these integrals. For the $B$-channel, we prove that the logarithmic part, yields:
\begin{equation}
I_{B(log)}^{(\Delta\tilde{U})(v\tilde{U})} + I_{B(log)}^{(\tilde{U}v)(\Delta\tilde{U})}= \frac{8\pi^2\mathbf{Z}[E_1]}{E_2(E_1 - E_2)} \cos\Big(2E_2\Delta u+\phi_2\Big) \Theta(E_1 - E_2)
\end{equation}
where we have defined $\phi_2=2(E_1-E_2)\;(a_2+4MD_0e^{\frac{-u_1+\beta_2}{4M}})$, with $a_2=v_2+4M$, $\beta_2=\Delta u+a_2$. In a similar manner, we find that 
\begin{equation}
I_{B(log)}^{(\Delta\tilde{U})(\tilde{U}v)} + I_{B(log)}^{(v\tilde{U})(\Delta\tilde{U})} = \frac{8\pi^2\mathbf{Z}[E_2]}{E_1(E_2 - E_1)} \cos\Big(2E_1\Delta u+\phi_1\Big)\Theta(E_2 - E_1)
\end{equation}
This term is exactly the previous one under the index swap $(1\leftrightarrow 2)$ for the coordinates $u,v$. We note that under the swap 1 $ \rightarrow $ 2, it is $\Delta u \rightarrow-\Delta u$ and $\beta_1 $ $   \rightarrow $ $ -\beta_2 $. These are summed up in the second-order coherence function: 
\begin{equation}
    C^{(2)}_{mem}(t_1, E_1; t_2, E_2) = \dfrac{E_1}{(E_1-E_2)}\; \frac{\mathbf{Z}[2E_1]}{\mathbf{Z}[E_1]\mathbf{Z}[E_2]} \;\Theta(E_1-E_2) \cos(2E_1\Delta u+\phi_2)\;\;+(1\leftrightarrow 2)
\end{equation}
where  $(1\leftrightarrow 2)$ indicates the same terms with swapping indices on the energies and the coordinates $u,v$.

For the Exponential-Integral terms, the corresponding second-order coherence function is:
\begin{equation}
    C^{(2)}_{corr}(t_1, E_1; t_2, E_2) = \dfrac{\mathbf{Z}[2E_1]}{\mathbf{Z}[E_1]\mathbf{Z}[E_2]}   \;\Theta(E_2-E_1) \cos(2E_1\Delta u) \;\;+(1\leftrightarrow 2).
\end{equation}

\begin{figure}[h!]
    \centering
    \includegraphics[width=0.45\textwidth]{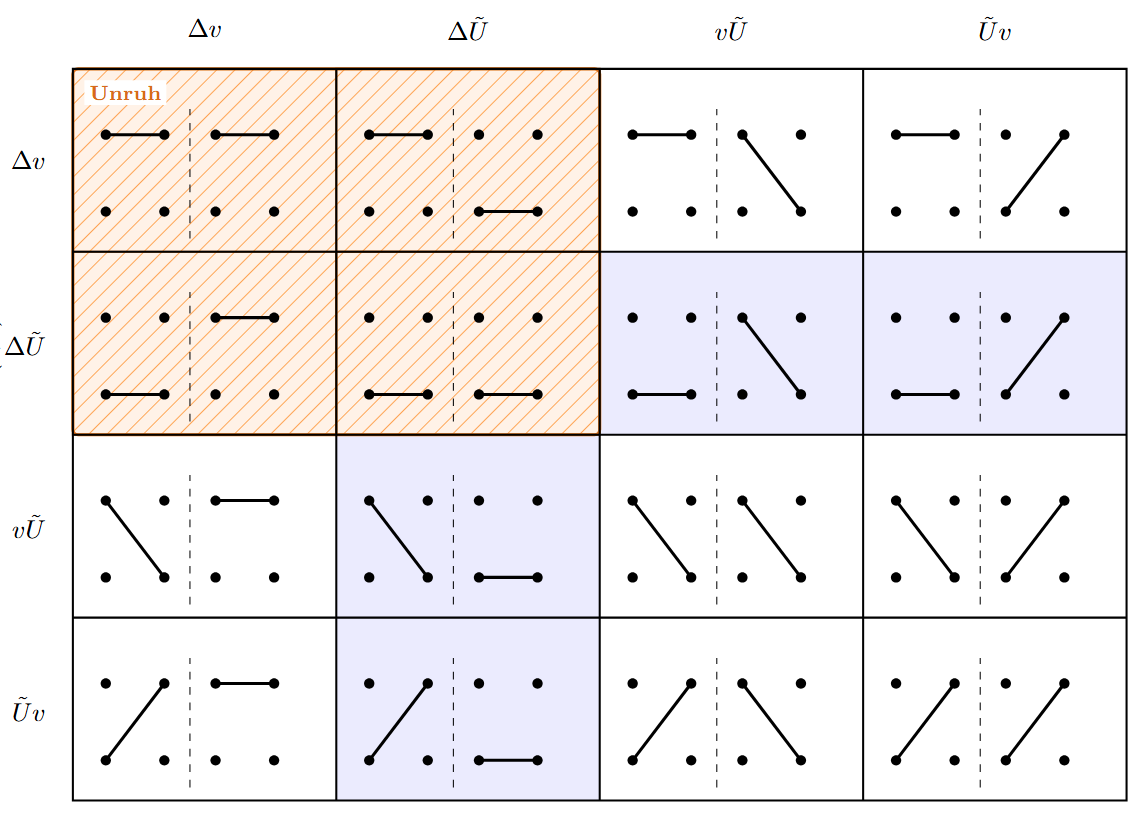}
     \raisebox{0.05cm}[0pt][0pt]{\includegraphics[width=0.455\textwidth]{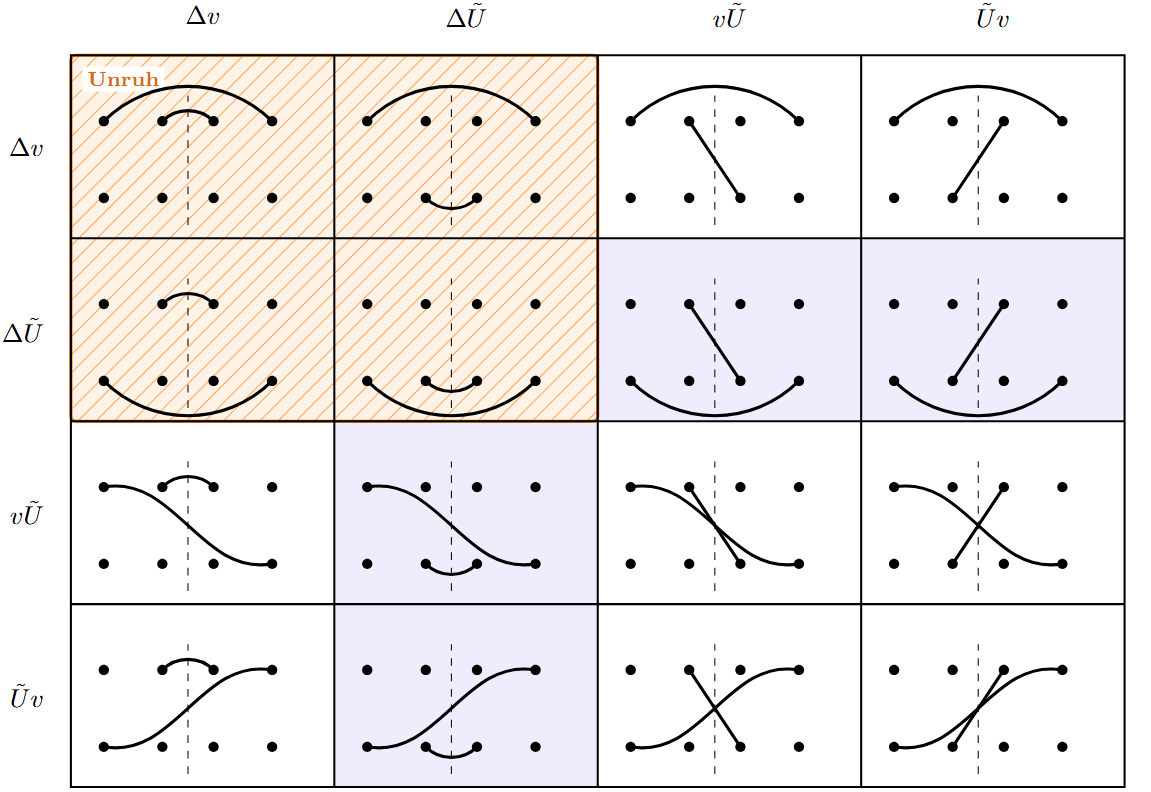}}
    \caption{4x4-Table of the representation of the integral contributions. \textbf{Left:}A-channel. \textbf{Right:}B-channel. The orange 2x2 block corresponds to the Unruh terms. The blue shaded areas correspond to integrals containing at least one $\Delta\tilde{U}$ term. These are the contributors of the $C_{mem}+C_{corr}$ correlation terms. }
\end{figure}


\begin{thebibliography}{}


 
\bibitem{Hawk1} S. W. Hawking, {\em Particle Creation by Black Holes}, Comm. Math. Phys. 43, 19 (1975).

\bibitem{Wald1} R.M. Wald, {\em On Particle Creation by Black Holes}, Comm. Math. Phys. 45, 9 (1975).


\bibitem{Page93} D. N. Page, {\em Information in Black Hole Radiation}, Phys. Rev. Lett. 71,   3743 (1993)

 \bibitem{AMPS} A. Almheiri, D. Marolf, J. Polchinski, and J. Sully, {\em Black Holes: Complementarity or Firewalls?}, 
JHEP 02, 062 (2013).

\bibitem{Mathur} S. D. Mathur, {\em The Information Paradox: A Pedagogical Introduction},
Class. Quant. Grav. 26, 224001 (2009).



\bibitem{Hawk76} S. W. Hawking,
{\em Breakdown of Predictability in Gravitational Collapse},
Phys. Rev. D14, 2460 (1976).

\bibitem{UnWa} W. G. Unruh and R. M. Wald, {\em Information Loss}, 	
	Rep.  Prog.  Phys. 80,   092002 (2017).


\bibitem{HBT} R. Hanbury Brown and R. Q. Twiss,
{\em Correlation between Photons in Two Coherent Beams of Light},
Nature 177, 27 (1956).

\bibitem{QOptics} L. Mandel and E. Wolf, 
{\em Optical Coherence and Quantum Optics} 
(Cambridge University Press, 1995).

\bibitem{NaBl} Y. V. Nazarov and Y. M. Blanter, {\em Quantum Transport: Introduction to Nanoscience} (Cambridge University Press, 2009).

\bibitem{BDZ} I. Bloch, J. Dalibard, and W. Zwerger,
{\em Many-Body Physics with Ultracold Gases}, Rev. Mod. Phys. 80, 885 (2008).
 

\bibitem{AnSav20} C. Anastopoulos and N. Savvidou, {\em Multi-Time Measurements in Hawking Radiation: Information at Higher-Order Correlations}, Class. Quant. Grav. 37, 025015 (2020).




\bibitem{QTP4} 
 C. Anastopoulos, B. L. Hu, and K. Savvidou, {\em Quantum Field Theory Based Quantum Information: Measurements and Correlations}, Ann. Phys. 450,  169239 (2023). 



\bibitem{ctp1}	J. S. Schwinger, {\em Brownian Motion of a Quantum Oscillator}, J. Math. Phys. 2, 407 (1961).

\bibitem{ctp2} L. V. Keldysh, {\em Diagram Technique for Nonequilibrium Processes}, Zh. Eksp. Teor. Fiz. 47, 1515 (1964).

\bibitem{ctp3} E. A. Calzetta and B. L. Hu,
{\em Nonequilibrium Quantum Field Theory}
(Cambridge University Press, 2008).


\bibitem{Glauber1} 	R. J. Glauber, {\em The Quantum Theory of Optical Coherence}, Phys. Rev. 130, 2529 (1963).

 \bibitem{Glauber2} 	R. J. Glauber, {\em Coherent and Incoherent States of the Radiation Field}, Phys. Rev. 131, 2766 (1963).

\bibitem{Page2} D. Page,  
{\em Particle Emission Rates from a Black Hole}, Phys. Rev. D13, 198 (1976).

  
\bibitem{HeKr} K. E. Hellwig and K. Kraus, {\em Formal Description of Measurements in Local Quantum Field Theory}, Phys.
Rev. D1, 566 (1970).

\bibitem{Sorkin} R. Sorkin, {\em Impossible Measurements on Quantum Fields}, in “Directions in General Relativity”, eds. B. L. Hu and T. A. Jacobson (Cambridge University Press, 1993).


 \bibitem{QTP1}	C. Anastopoulos and N. Savvidou, {\em Time-of-Arrival Probabilities for General Particle Detectors}, Phys. Rev. A86, 012111 (2012).



 
\bibitem{QTPdet}
 C. Anastopoulos and N. Savvidou, {\em Measurements on Relativistic Quantum Fields: I. Probability Assignment}, arXiv:1509.01837.

 \bibitem{OkOz} K. Okamura and M. Ozawa, {\em Measurement Theory in Local Quantum Physics}, J. Math. Phys. 57, 015209
(2015).


\bibitem{QTP3}	C. Anastopoulos and N. Savvidou, {\em Time of Arrival and Localization of Relativistic Particles}, J. Math. Phys. 60, 032301 (2019).


  
\bibitem{FeVe} C. J. Fewster and R. Verch, {\em Quantum Fields and Local Measurements}, Comm. Math. Phys. 378, 851 (2020).

 





\bibitem{GGM22}  J. Polo-Gómez, L. J. Garay, L. J. and E. Martín-Martínez,  {\em A Detector-Based Measurement Theory for Quantum Field Theory},  	Phys. Rev. D 105, 065003 (2022).

 

\bibitem{PTM} T. R. Perche, J. Polo-Gómez, B. de S. L. Torres, and E. Martín-Martínez, {\em Particle Detectors from Localized Quantum Field Theories},  	Phys. Rev. D109, 045013 (2024).

  

 
 
 
\bibitem{Unruh76} W. G. Unruh, {\em Notes on Black Hole Evaporation}, Phys. Rev. D14, 870 (1976).


\bibitem{BiDa} N. D. Birrell and P. Davies, {\em Quantum Fields in Curved Space} (Cambridge University Press, 1982).

\bibitem{Dewitt}B. S. DeWitt, {\em Quantum Gravity: the New Synthesis} in
General Relativity: An Einstein Centenary Survey, ed. by S. W. Hawking
and W. Israel (Cambridge University Press, Cambridge, 1979), p.
680.


\bibitem{AnSav11}	C. Anastopoulos and N. Savvidou, {\em Coherences of Accelerated Detectors and the Local Character of the Unruh Effect}, J. Math. Phys. 53, 012107 (2012).





\bibitem{AnSav19} C. Anastopoulos and N. Savvidou, {\em Multi-Time Measurements in Hawking Radiation: Information at Higher-Order Correlations}, Class. Quant. Grav. 37, 025015 (2020).

\bibitem{Park75} L. Parker, {\em Probability Distribution of Particles Created by a Black Hole}, Phys. Rev. D12, 1519 (1975).

\bibitem{QFTCSWald} R. M. Wald, {\em Quantum Field Theory in Curved Spacetime and Black Hole Thermodynamics} (University of Chicago Press, 1994).

\bibitem{Klau70} J. Klauder, {\em Exponential Hilbert Space: Fock Space Revisited }, J. Math Phys. 11, 609 (1970).

\bibitem{Dav}  E. B.  Davies,  {\em Quantum Theory of Open Systems} (Academic Press, London 1976).

\bibitem{BrePe}H. P. Breuer and F. P. Petruccione, {\em The Theory of Open Quantum Systems} (Oxford University Press, 2007).


\bibitem{PaZu93} H. P. Paz and W. H. Zurek, {\em Environment-Induced Decoherence, Classicality, and Consistency of Quantum Histories}  Phys. Rev. D48, 2728 (1993).

\bibitem{VA17} I De Vega and D. Alonso, {\em Dynamics of non-Markovian Open Quantum Systems}, Rev. Mod. Phys. 89, 015001 (2017).

\bibitem{Corless02} R. M. Corless and D. J. Jeffrey, {\em The Wright $\omega$ Function}, Lect. Notes Comput. Sci. 2385, 76 (2002). 


\end{thebibliography}
\end{document}